\documentclass[twocolumn]{aastex631}

\usepackage{amssymb,amsmath,verbatim,mathtools,needspace,enumitem,etoolbox,graphicx,physics,microtype,afterpage,xspace,tabularx,lmodern,multirow}

\newcommand{\jhu}{William H. Miller III Department of Physics and Astronomy, Johns Hopkins University, Baltimore, Maryland 21218, USA}

\newcommand{\nh}{\mbox{{\sc \small NewHorizon}}~}
\newcommand{\hagn}{\mbox{{\sc \small Horizon-AGN}}~}

\shorttitle{SMBHs grow in hierarchically merging NSCs}
\shortauthors{K.~Kritos et al.}

\begin{document}

\title{Supermassive black hole growth in hierarchically merging nuclear star clusters}

\author[0000-0002-0212-3472]{Konstantinos Kritos}
\correspondingauthor{Konstantinos Kritos}
\email{kkritos1@jhu.edu}
\affiliation{\jhu}

\author[0000-0002-2850-0192]{Ricarda S. Beckmann}
\affiliation{Institute of Astronomy and Kavli Institute for Cosmology, University of Cambridge, Madingley Road, Cambridge, CB3 0HA, UK}
\affiliation{Institute for Astronomy, University of Edinburgh, Royal Observatory, Edinburgh EH9 3HJ, UK}

\author[0000-0002-1566-8148]{Joseph Silk}
\affiliation{\jhu}
\affiliation{Institut d’Astrophysique de Paris, UMR 7095 CNRS and UPMC, Sorbonne Universit$\acute{e}$, F-75014 Paris, France}
\affiliation{Department of Physics, Beecroft Institute for Particle Astrophysics and Cosmology, University of Oxford, Oxford OX1 3RH, United Kingdom}

\author[0000-0003-0751-5130]{Emanuele Berti}
\affiliation{\jhu}

\author[0000-0002-9104-1734]{Sophia Yi}
\affiliation{\jhu}

\author[0000-0002-3216-1322]{Marta Volonteri}
\affiliation{Institut d’Astrophysique de Paris, CNRS, Sorbonne Université, UMR7095, 98bis bd Arago, 75014 Paris, France}

\author[0000-0003-0225-6387]{Yohan Dubois}
\affiliation{Institut d’Astrophysique de Paris, CNRS, Sorbonne Université, UMR7095, 98bis bd Arago, 75014 Paris, France}

\author[0000-0002-8140-0422]{Julien Devriendt}
\affiliation{Department of Physics, University of Oxford, Keble Road, Oxford, OX1 3RH, UK}

\begin{abstract}
Supermassive black holes are prevalent at the centers of massive galaxies, and their masses scale with galaxy properties, increasing evidence suggesting that these trends continue to low stellar masses.
Seeds are needed for supermassive black holes, especially at the highest redshifts explored by the James Webb Space Telescope. We study the hierarchical merging of galaxies via cosmological merger trees and argue that the seeds of supermassive black holes formed in nuclear star clusters via stellar black hole mergers at early epochs. Observable tracers include intermediate-mass black holes, nuclear star clusters, and early gas accretion in host dwarf galaxies, along with a potentially detectable stochastic gravitational wave background, ejection of intermediate and supermassive black holes, and consequences of a significant population of tidal disruption events and extreme-mass ratio inspirals.
\end{abstract}

\date{\today}

\section{Introduction}
\label{sec:Introduction}

Supermassive black hole (SMBH) formation is an outstanding problem in contemporary cosmology and astrophysics.
SMBHs often coexist with nuclear star clusters (NSCs), the densest known stellar systems in the Universe, found in galactic nuclei~\citep{Neumayer:2020gno}.
Moreover, scaling relations between the SMBH's and NSC's mass have been suggested~\citep{2016MNRAS.457.2122G}, which may imply a coevolution history, and the NSC might have contributed to the formation of black hole (BH) seeds that further grow into SMBHs.

Thanks to its infrared capabilities, the James Webb Space Telescope (JWST) has revolutionized our understanding of the high-redshift Universe, and a large population of distant SMBHs has been revealed~\citep{2024ApJ...964...39G,Maiolino:2023bpi,Maiolino:2023zdu,2024arXiv240303872J,Kovacs:2024zfh}. It is generally believed that SMBHs have grown out of smaller BH seeds that may be light or massive. The former designates seeds of stellar origin with masses up to $100\,M_\odot$, and the latter refers to seeds of intermediate black hole mass (IMBH) in the $\sim 10^3$--$10^5\,M_\odot$ range regime~\citep{2020ARA&A..58..257G}.
The existence of stellar-mass BHs is observationally well established, and that of massive seeds has theoretical backing from Population III studies motivated by tentative detections of spectroscopic signatures from a metal-free stellar population at high redshift, including strong HHEII emission, extremely blue UV continua, and enhanced [N/O]~\citep{2024MNRAS.534..290L}.
The observation of SMBHs at high redshift ($z>6$) implies that seed formation and growth must have occurred rapidly within a few hundred $\rm Myr$ (at least for a subset of the observed massive BHs).
Moreover, a number of strongly-lensed magnified massive (masses $\sim10^6\,M_\odot$) compact (radii $<1\,\rm pc$) young star clusters have been detected by JWST at $z\sim10$~\citep{2024Natur.632..513A}, and these could be the sites of massive BH seed formation.
According to simulations, the assembly of such massive star clusters is much more efficient at $z>7$~\citep{2024arXiv241100670M}.

Theoretical works that study the growth of SMBHs in the cosmological context rely either on expensive cosmological simulations or semianalytic methods, which are faster to implement. 
A number of previous works have studied the formation and evolution of massive BHs in the cosmological context~\citep{DiMatteo:2007sq,Bonoli:2008xe,Natarajan:2011dv,Dubois:2014yya,Sijacki:2014yfa,2014MNRAS.440.1590D,Tremmel:2015rra,2020MNRAS.498.2219V,Izquierdo-Villalba:2021prf,2023A&A...676A...2D,2023A&A...673A.120D,Bhowmick:2023prn,2024arXiv241202374D,2024arXiv241007856L,Bhowmick:2024hfz}.
Cosmological simulations follow the three-dimensional coupled evolution of dark matter, gas, stars, and massive BHs.
Due to the many orders of magnitude of length scale required to be simulated, simulations are typically too expensive to capture scales below $100\,\rm pc$ over long periods of time. 
Therefore, simulating the formation process of massive BH seeds is nontrivial because they occur on spatial and time scales much smaller than can be resolved. 
Consequently, these cosmological simulations must rely on extrapolated models to place seed BHs in galaxies.
At the same time, gravitational runaway processes occur in the cores of dense star clusters at the subparsec scale and operate on time scales of a few $\rm Myr$. 
Therefore, in this work, we simulate the growth of these light BH seeds assuming that they are initially formed by the usual astrophysical process of massive star evolution. 

In previous work, \cite{Kritos:2024upo} developed {\sc Nuce}, a code that self-consistently evolves NSCs, and studied the formation of SMBHs in NSCs, simulating the seeding process. The only true seeds that must be put {\it by hand} in NSCs are stellar-mass BHs with mass $\sim10\,M_\odot$ that form after the evolution of massive stars. A number of works have emphasized the importance of a dense stellar cluster environment in rapidly forming massive BH seeds through mergers and accretion~\citep{1987ApJ...321..199Q,PortegiesZwart:2002iks,Omukai:2008wv,Devecchi:2010ps,Devecchi:2012nw,Alexander:2014noa,Mapelli:2016vca,Antonini:2018auk,Reinoso:2018bfv,Natarajan:2020avl,Vergara:2022trz,Kritos:2022non,Rantala:2024crf,Mehta:2024jgy,Prole:2024koa,Partmann:2024ees,2024arXiv240918605D}. Alternative scenarios also exist, such as primordial BHs~\citep{Hooper:2023nnl,Ziparo:2024nwh} and the growth of Population III stars~\citep{Freese:2010re,Cammelli:2024uhh}.

In this paper we present a semi-analytic model for the evolution of NSCs and their central massive BHs based on galaxy merger trees extracted from the \nh simulation~\citep{2021A&A...651A.109D};
see e.g. \cite{Polkas:2023isn} for another semianalytic model of NSCs.
In Sec.~\ref{sec:Methods}, we present our methodology for coupling the {\sc Nuce} code with those merger trees; in Sec.~\ref{sec:Results}, we show our results on generation of massive BH and coexisting NSC populations; in Sec.~\ref{sec:Caveats}, we  suggest that associated stellar tidal disruption events (TDEs) in NSCs will be an inevitable corollary of our model and discuss  more generally the caveats of our study, and 
in Sec.~\ref{sec:Conclusions} we present our conclusions.{\footnote{
In this manuscript, we use {\it massive BH} to refer to any BH with a mass $>100\,M_\odot$. However, since we focus on a mass range with more than 10 orders of magnitude, it is common practice to further categorize massive BHs into IMBHs in the range $100\,M_\odot$--$10^6\,M_\odot$ and SMBHs with masses $>10^6\,M_\odot$.
}}
 
\section{Methods}
\label{sec:Methods}

In this section we present our assumptions and methodology.
In Sec.~\ref{sec:Merger_trees} we introduce the \nh simulation from which we extract merger trees. In Sec.~\ref{sec:NSC_initial_conditions} we discuss NSC initial conditions. In Sec.~\ref{sec:Massive_BH_seeds} we describe the formation of massive BH seeds in NSCs. Finally, in Sec.~\ref{sec:Major_merger_episodes} we present our algorithm for treating NSC-NSC coalescence and massive BH binary formation and merger during major galaxy-galaxy mergers.

\subsection{Merger trees}
\label{sec:Merger_trees}

In this paper we build our model of NSCs and IMBHs on galaxy evolution histories extracted from the \nh simulation. \nh has been presented in detail in \cite{2021A&A...651A.109D}, so we only briefly recap its relevant features here.

\nh is a high-resolution resimulation of a sphere of radius $\sim16 \,\rm Mpc$ of the \hagn simulation. \nh was performed with {\sc ramses}~\citep{Teyssier:2001cp} and uses a $\Lambda$CDM cosmology consistent with WMAP-7 data~\citep{2011ApJS..192...18K}. The simulation reached a maximum resolution of 34 comoving pc using a quasi-Lagrangian and super-Lagrangian refinement scheme. Gas follows an ideal monoatomic equation of state with an adiabatic index of $\gamma_{\rm ad}=5/3$. Cooling is modeled down to $10^4\, \rm K$ using cooling curves from~\citet{1993ApJS...88..253S} assuming equilibrium chemistry. After redshift $z_{\rm reion} = 10$, heating from a uniform UV background is included following~\citet{Haardt:1995bw}. Star formation takes place following a Schmidt relation~\citep{2017MNRAS.466.4826K,Trebitsch:2017nji,2020MNRAS.494.3453T} in cells with a gas number density above $n_0=10\, \rm H\, cm^{-3}$ using a star formation efficiency that depends on the properties of the interstellar medium such as the Mach number and virial parameters. Stars have a stellar mass resolution of $ 1.3\times 10^4 \,M_\odot$ and are assumed to have a Chabrier~\citep{Chabrier:2004vw} initial mass function with cutoffs at $0.1\,M_\odot$ and $150\,M_\odot$. Stellar feedback separately tracks the momentum and energy-conserving phase of the explosion~\citep{2015MNRAS.451.2900K}. \nh contained on-the-fly evolution for BH formation and evolution, but as their properties are not included in this analysis, we omit discussing this further. Interested readers are referred to~\cite{Beckmann:2022siq}.

Halos and galaxies were identified in the simulation using the HOP structure finding algorithm~\citep{Aubert:2004mu}, using an overdensity of $\delta_{\rm halos} = 80$ (for halos) and  $\delta_{\rm galaxies} = 160$ for galaxies. Uncontaminated halos only contain high-resolution dark-matter particles within their virial radius. 
Our sample of target galaxies consists of all central galaxies with a stellar mass above $6.5 \times 10^ 5 \, M_\odot$ (equivalent to at least 50 star particles) that are located within the central 0.1 virial radii of a dark-matter halo at $z=0.25$.

\begin{figure}
    \centering
    \includegraphics[width=\linewidth]{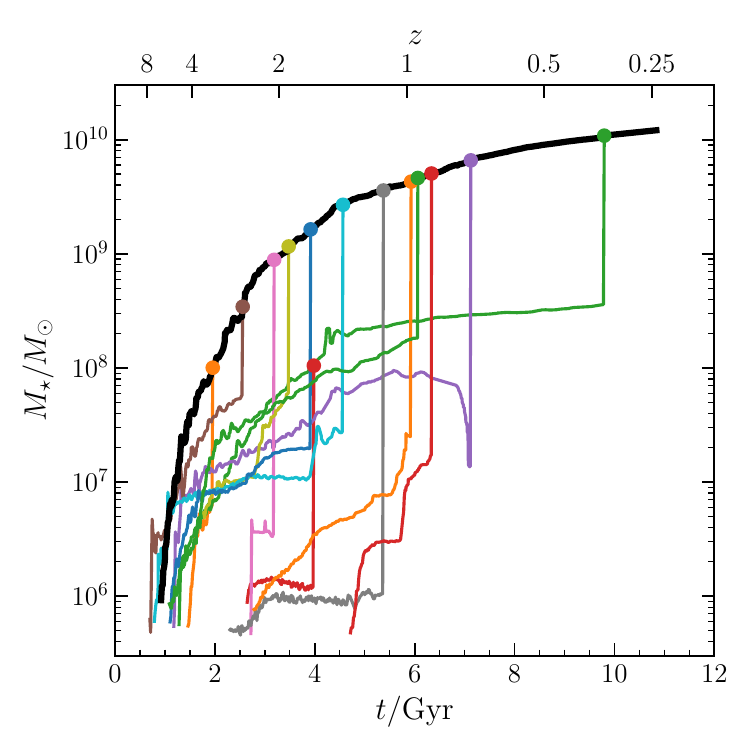}
    \caption{Example galaxy merger tree showing the galaxy stellar mass evolution of the main progenitor (black), as well as all other progenitors (colored lines) that reach a mass of  $M_\star > 10^7 \rm \ M_\odot$ before the galaxy merger. Colored markers mark galaxy mergers.}
    \label{fig:merger-tree-example}
\end{figure}

From this sample of galaxies, we construct merger trees across cosmic time (see Fig.~\ref{fig:merger-tree-example} for an example) that show the stellar mass evolution history of the galaxy's progenitors across cosmic time. Merger trees are constructed by analyzing a set of 997 simulation snapshots in the redshift range $z=10$--$0.25$. We only retain galaxies that attain a mass of at least $10^7\, M_\odot$ in stellar mass before the merger and discard all others. 

\subsection{NSC initial conditions}
\label{sec:NSC_initial_conditions}

In our NSC paradigm, SMBH seeds form from repeated merger processes and accretion in the dense centers of protogalaxies. 
We assume that each protogalaxy forms an NSC in a star formation burst when its stellar mass reaches a critical value, $M_{\rm cr}$. 
The initial stellar mass $M_{\star,0}$ contained in the NSC is then taken to be a random fraction of $M_{\rm cr}$, denoted by $f_\star\in[0, f_{\star,\rm max}]$, and we write $M_{\star,0} = f_\star M_{\rm cr}$. 

Assuming a Kroupa initial mass function in the range $0.08M_\odot$ to $150M_\odot$ with the mean parameters from~\cite{Kroupa:2002ky} (which we assume to be universal), the initial average stellar mass is $\overline{m}_{\star,0}=0.6M_\odot$.
The initial number of stars is then computed as $N_{\star,0} = M_{\star,0} / \overline{m}_{\star,0}$ which is equivalent to $N_{\star,0} = f_{\star} M_{\rm cr}/\overline{m}_{\star,0}$.

An amount of gas is also added initially into the NSC such that a random fraction $f_{\rm g}\in[0, f_{\rm g,max}]$ of the cluster’s mass is in the form of gas that did not fragment into stars. 
Thus, if $M_{\rm cl,0} = M_{\star,0} + M_{\rm g,0}$ is the total initial mass of the NSC, then $M_{\rm g,0} = f_{\rm g} M_{\rm cl,0}$ is the initial amount of gas in the system not turned into stars assumed to be in the form of ionized hydrogen with a sound speed of $c_{\rm s}=10\, \rm km\, s^{-1}$. 
We compute the initial gas mass as $M_{\rm g,0} = M_{\star,0} f_{\rm g} / (1 - f_{\rm g})$ and the star formation efficiency is given by $\eta_\star=1 - f_{\rm g}$.
The gas is completely removed from the NSC with an exponential time dependence on a time scale $\tau_{\rm ge}$, which is a free parameter of our model~\citep{Baumgardt:2007av}.

The prior on the initial half-mass radius $r_{\rm h,0}$ is drawn from a log-uniform distribution from $0.1\,\rm pc$ to $10\,\rm pc$. 
This choice is supported by young massive star cluster properties in the local Universe~\citep{2010ARA&A..48..431P} and by recent observations of high-redshift young star clusters with effective radii on parsec and sub-parsec scales~\citep{2023ApJ...945...53V,2024Natur.632..513A}.
Moreover, according to~\cite{2010MNRAS.409.1057D}, NSCs may form in protogalaxies with $r_{\rm h,0}<0.5\,\rm pc$.
Since most NSCs observed in the local Universe have effective radii of a few $\rm pc$~\citep{Neumayer:2020gno}, and by H\'enon’s principle, energy generation in their core causes them to expand during balanced evolution, then these systems must have been more compact in the past when they formed. 
The diversity of radii in present-day NSCs indicates a variety of initial radii and possibly different evolutionary conditions. 

We evolve NSCs and simulate the formation of SMBH seeds using the numerical code {\sc Nuce} (for ``Nuclear cluster evolution''), a semianalytic model that relies on H\'enon’s principle, described in~\cite{Kritos:2024upo}. 

\subsection{Massive BH seeds}
\label{sec:Massive_BH_seeds}

In this subsection we describe the BH growth channels considered in our model (Sec.~\ref{sec:BH_growth_channels}). Then we show the evolution of the mass and spin of the seeds (Sec.~\ref{sec:Time_evolution_of_MBH_and_xBH}), and finally we present the dependence of seed formation on the NSC initial conditions (Sec.~\ref{sec:Mapping_BH_masses_in_the_NSC_parameter_space}).

\subsubsection{BH growth channels}
\label{sec:BH_growth_channels}

Our NSC model {\sc Nuce} implements the following seeding channels:
\begin{itemize}
\item {\bf Stellar collisions}: Successive physical collisions among stars in the cluster lead to the formation of a very massive star ($>150M_\odot$), which then directly collapses into a massive BH seed. This runaway process initiates provided that $0.2\tau_{\rm rh,0}<\tau_{\rm se}=3\,\rm Myr$, where $\tau_{\rm rh,0}$ is the initial half-mass relaxation time scale and $\tau_{\rm se}$ is the massive star evolution time scale. The mass of the massive object is a fraction $\sim10^{-3}$ of the initial stellar mass~\citep{PortegiesZwart:2002iks,Rasio:2003sz}. Since runaway stellar collisions rapidly occur in the first few $\rm Myr$ if the previous condition is met~\citep{Fujii:2024uon}, we do not simulate them, and rather assume that a massive BH seed has formed in the cluster with mass $M_{\rm BH}=10^{-3}N_{\star,0}\overline{m}_{\star,0}$. This BH may then grow further through the following channels. If the runaway condition is not met, then the seed BH mass is set to be $10\,M_\odot$.
    \item {\bf Gas accretion}: We assume that gas is accreted onto BHs at the Bondi rate~\citep{1952MNRAS.112..195B} capped to the Eddington limit and that accretion operates only for as long as there is remaining gas in the cluster. For the gas density, we take $\rho_{\rm g}=M_{\rm g}/(2r_{\rm h}^3)$ and we assume a sound speed of $c_{\rm s}=10\,\rm km\,s^{-1}$~\citep{Kritos:2024upo}. We recall that the gas is removed exponentially with decay time scale $\tau_{\rm ge}$~\citep{Baumgardt:2007av}.
    \item {\bf BH mergers}: Binary BHs form after the core collapse of the BH subsystem, and if a binary does not receive an interaction recoil that exceeds the escape velocity $v_{\rm esc}$, it then merges within the host cluster as a consequence of hardening. The binary is ejected if the GW kick $v_{\rm GW}$ imparted to the post-merger remnant is such that $v_{\rm GW}>v_{\rm esc}$; otherwise, it is retained in the cluster and it may merge with another BH. These repeated BH mergers lead to BH growth.
    \item {\bf Star consumption}: The contribution of repeated stellar consumption events becomes nonvanishing in our model only after the BH subsystem has evaporated, within approximately 10 initial half-mass relaxation times~\citep{Breen:2013vla}. We compute the full-loss-cone stellar consumption rate implementing Eq.~(6.15) from~\cite{2013degn.book.....M} assuming a Bahcall-Wolf cusp. The loss-cone radius is set to double the tidal radius to account for tidal captures~\citep{Rizzuto:2022fdp}. We assume that a mass of $0.05\overline{m}_{\star,0}$ is accreted.
\end{itemize}

\begin{figure}
    \centering
    \includegraphics[width=\linewidth]{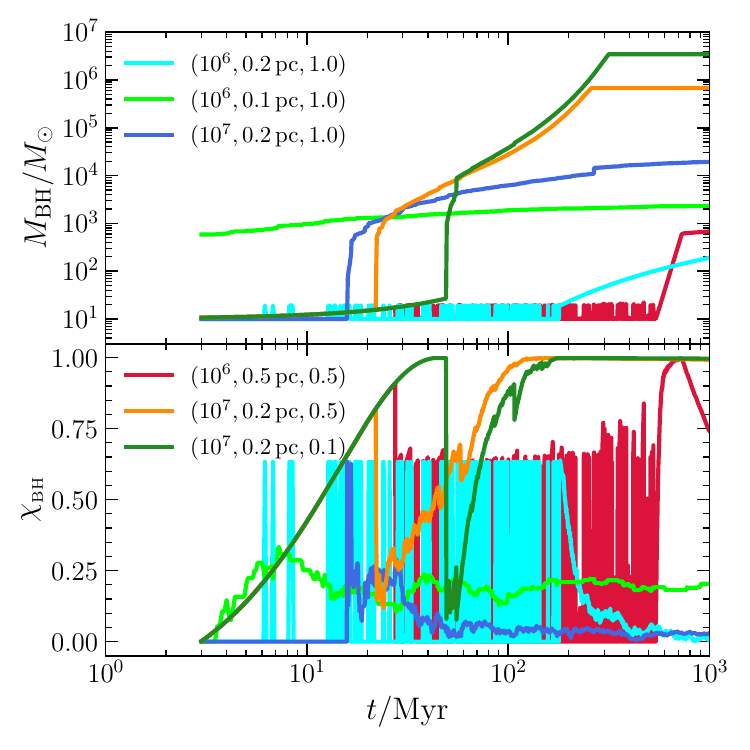}
    \caption{The time evolution of BH mass ($M_{\rm BH}$, upper panel) and dimensionless spin ($\chi_{\rm BH}$, lower panel). Different colors correspond to various initial conditions for $(N_{\star,0}, r_{\rm h,0}, \eta_\star)$, as shown in the legends.}
    \label{fig:MBH-xBH-t}
\end{figure}

\subsubsection{Time evolution of $M_{\rm BH}$ and $\chi_{\rm BH}$}
\label{sec:Time_evolution_of_MBH_and_xBH}

In this subsection we test the massive BH growth model in NSCs without using the merger trees described in Sec.~\ref{sec:Merger_trees}.
In Fig.~\ref{fig:MBH-xBH-t} we show the time evolution of the mass ($M_{\rm BH}$) and spin ($\chi_{\rm BH}$) of massive BH seeds when we consider different initial conditions for $(N_{\star,0}, r_{\rm h,0}, \eta_\star)$, i.e., the initial number of stars, half-mass radius, and star formation efficiency, respectively.

We compute the time evolution of the BH spin from the relation $d\chi_{\rm BH}/dt=(d\chi_{\rm BH}/dM_{\rm BH})(dM_{\rm BH}/dt)$, where $d\chi_{\rm BH}/dM_{\rm BH}$ is computed by following the classic work of~\cite{Bardeen:1970zz}: see Eq.~(1) in~\cite{Kritos:2024kpn}. The factor $dM_{\rm BH}/dt$ contains contributions from gas accretion and consumption of stars. 
We assume that accretion (of gas or stars) is coherent along the plane orthogonal to the direction of the BH spin, and we randomize the direction of the accretion (prograde or retrograde) with a probability $50\%$.

Below we describe in detail the evolution of $M_{\rm BH}$ and $\chi_{\rm BH}$ in the specific cases shown in Fig.~\ref{fig:MBH-xBH-t}. 

\noindent
\paragraph{The case $(10^6, 0.2\,{\rm pc}, 1.0)$.} The cluster does not satisfy the conditions for runaway stellar collisions or repeated BH mergers to assemble an IMBH. Second-generation BHs ($\simeq20\,M_\odot$) form and merge with other first-generation BHs ($10\,M_\odot$). Still, the third-generation remnants of these mergers are ejected promptly due to the cluster's relatively small escape speed, which is $\simeq185\,\rm km\, s^{-1}$ initially.
In particular, since second-generation merger products of comparable mass BHs have a spin of $\sim0.7$~\citep{Baibhav:2021qzw} and first-generation BHs are assumed to be nonspinning, the probability of retaining the third-generation product is about $4\%$.
This repeated formation of second-generation BHs and subsequent ejections results in the oscillatory behavior seen in the mass and spin evolution plots.
Once 90\% of the BHs are ejected at $\simeq200\,\rm Myr$, repeated consumption of stars initiates and slowly leads to the formation of a $\approx200\,M_\odot$ massive BH (within $1\,\rm Gyr$). At the same time, the spin asymptotes to zero.

\noindent
\paragraph{The case $(10^6, 0.1\,{\rm pc}, 1.0)$.} The cluster has the required initial conditions to undergo runaway stellar collisions, and a $\approx600\,M_\odot$ massive BH forms by $t=3\,\rm Myr$. We assume that the very massive stellar progenitor directly collapses into a BH without mass loss, and we (arbitrarily) set the spin of the direct collapse BH to zero. By $t=1\,\rm Gyr$, the IMBH further grows to $\approx2000\,M_\odot$ by merging with other BHs and consuming stars, while its spin hovers around $\chi_{\rm BH}\sim0.2$ within the simulated time window.

\noindent
\paragraph{The case $(10^7, 0.2\,{\rm pc}, 1.0)$.} The cluster is massive and compact enough for the initial escape velocity to exceed $300\,\rm km\,s^{-1}$, which is the threshold for hierarchical mergers to happen. A $10^3\,M_\odot$ massive BH is formed within a few tens of $\rm Myr$  and grows to $\approx2\times10^4\,M_\odot$ by $t=1\,\rm Gyr$. The spin of this BH increases to $\sim0.7$ after the first merger and drops below $\chi_{\rm BH}\sim0.1$ by the end of the simulation as a result of the isotropic spin directions in merger events. 

\noindent
\paragraph{The case $(10^6,0.5\,{\rm pc}, 0.5)$.} As in the first case above, the probability of forming a massive BH from repeated BH mergers is small due to the insufficient escape velocity of the system. The system contains a lot of gas but the gas accretion rate is too low to grow the BH, because the mass-doubling time scale (based on the Bondi rate) is smaller than the gas expulsion time scale. Only after the initiation of stellar consumption at $\approx530\,\rm Myr$ (following the evaporation of the BH subsystem) does the gas accretion lead to BH growth, and spins it up to the maximal value. At $\approx700\,\rm Myr$ the remnant gas has been removed completely, and further growth proceeds through stellar consumption. During this final phase the BH spin drops, asymptoting to zero on the mass doubling time as a consequence of repeated consumption of stars.

\noindent
\paragraph{The cases $(10^7,0.2\,{\rm pc},0.5)$ and $(10^7,0.2\,{\rm pc}, 0.1)$.} We consider the same structural initial conditions as in the case $(10^7, 0.2\,{\rm pc}, 1.0)$ above, but we consider different values for the star formation efficiency. In both cases the seed BH grows by mergers and gas accretion that act simultaneously, but on different time scales. Depending on the amount of gas present in the system, the final BH reaches $\simeq7\times10^5\,M_\odot$ ($\eta_\star=0.5$) or $\simeq4\times10^6\,M_\odot$ ($\eta_\star=0.1$) within $300\,\rm Myr$. The BH starts to slowly grow by gas accretion in the first $\simeq50\,\rm Myr$ and spins up to the maximal value, but the accretion rate is enhanced once repeated BH mergers lead to BH masses $\sim10^3$--$10^4\,M_\odot$. During the hierarchical BH merger phase, the spin drops to low values $\sim0.1$ (see the dips in the evolution of the spin in the orange and green lines). Once gas accretion takes over the evolution of the BH, it steadily and rapidly grows into the SMBH regime; the spin magnitude grows back to the maximal value within a mass doubling time and remains at the maximal value, since we assume coherent gas accretion. We remind the reader that we have chosen to isotropically accrete stars and stellar-mass BHs onto the growing BH, while gas is accreted coherently.

\subsubsection{Mapping BH masses in the NSC parameter space}
\label{sec:Mapping_BH_masses_in_the_NSC_parameter_space}

\begin{figure*}
    \centering
    \includegraphics[width=\textwidth]{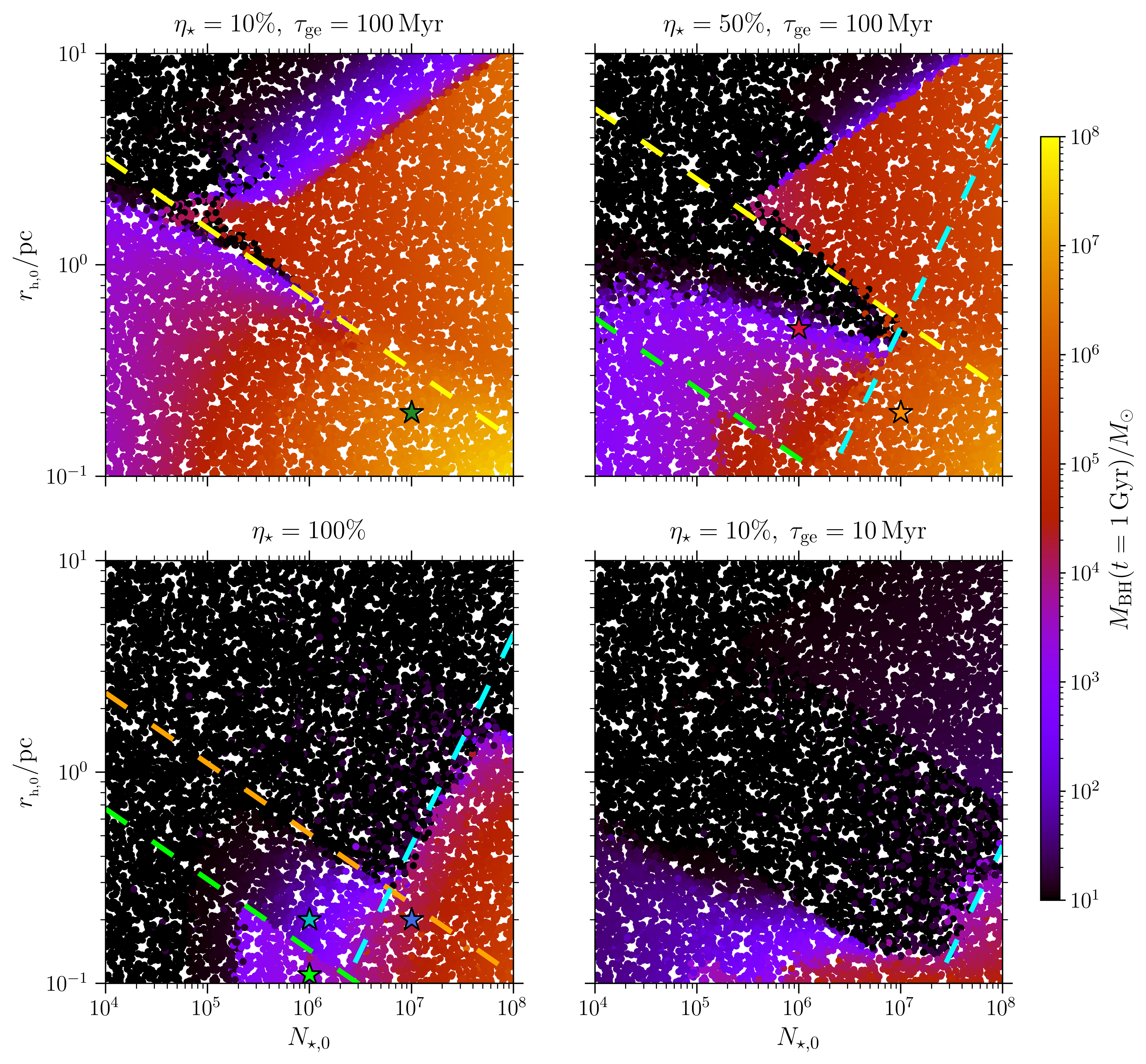}
    \caption{Masses of the BH seeds $M_{\rm BH}$ formed through the combined processes of successive mergers and gas accretion within $1\,\rm Gyr$ in NSCs as a function of the initial number of stars $N_{\star,0}$ and initial half-mass radius $r_{\rm h,0}$. Each point corresponds to an NSC simulation with {\sc Nuce}. A few pairs of the star formation efficiency ($\eta_\star$) and gas expulsion time scale ($\tau_{\rm ge}$) have been assumed and shown in the title of each panel. The colored stars in the panels represent the set of initial conditions for the simulations of Fig.~\ref{fig:MBH-xBH-t}.}
    \label{fig:seed_mass_dependence_on_initial_cluster_conditions}
\end{figure*}

In Fig.~\ref{fig:seed_mass_dependence_on_initial_cluster_conditions} we show the mass of the heavier BH formed within $1\,\rm Gyr$ for $10^4$ randomly chosen initial conditions in the ranges $10^4\le N_{\star,0}\le10^8$ and $0.1\le r_{\rm h,0}/{\rm pc}\le10$.

The boundaries of massive BH formation are easier to understand in the dry ($\eta_\star=100\%$) case without gas accretion. 
Clusters with initial conditions below the lime dashed line of Fig.~\ref{fig:seed_mass_dependence_on_initial_cluster_conditions} satisfy the condition for runaway stellar collisions and form a massive BH within $3\,\rm Myr$.
Clusters with initial conditions below the orange-dashed line eject their stellar-mass BHs and undergo runaway stellar consumption. 
The normalization constant of the orange-dashed boundary depends on the time since the beginning of the simulation, and in all panels of Fig.~\ref{fig:seed_mass_dependence_on_initial_cluster_conditions}, we show the BH mass at $t=1\,\rm Gyr$. Since the BH subsystem evaporates within around 10 initial half-mass relaxation times~\citep{Breen:2013vla}, the condition $10\tau_{\rm rh,0}<t$ is required for runaway stellar consumption to have an effect. Otherwise, the BH subsystem dominates the core, and no massive BH forms unless repeated BH mergers occur. Thus, the parameter space of massive BHs increases with time.
If the cluster's initial conditions lie below the cyan dashed line -- corresponding to an initial escape velocity of $300\,\rm km\, s^{-1}$ -- then repeated BH mergers can form a massive BH. This result agrees with the findings of~\cite{Antonini:2018auk} for the threshold for massive BH formation via repeated mergers.

Now we discuss the rest of the panels of Fig.~\ref{fig:seed_mass_dependence_on_initial_cluster_conditions}. The presence of gas in the system boosts the growth of the massive BH, and the available parameter space expands as long as the gas remains in the system long enough, as we argue below.
In the region below the yellow lines in the upper panels of Fig.~\ref{fig:seed_mass_dependence_on_initial_cluster_conditions} BH interactions become efficient, and the binary formation rate is much smaller than $100\,\rm Myr$. The growing BH merges with other stellar-mass BHs in the core and eventually gets ejected due to GW recoil, so it does not have time to grow. Only after most BHs have been ejected is the successive consumption of stars turned on, which contributes to the formation of massive BHs.
In the high $r_{\rm h,0}$ region, the gas density is not high enough, hence accretion at the Bondi rate becomes insufficient to grow the BH in the available $100\,\rm Myr$ before it is removed, because we assume the gas density to be $\rho_{\rm g}\simeq M_{\rm g}/(2r_{\rm h}^3)$~\citep{Kritos:2024upo}.

In the case $\tau_{\rm ge}=10\,\rm Myr$ (lower-right panel of Fig.~\ref{fig:seed_mass_dependence_on_initial_cluster_conditions}) the gas is removed faster, and the cluster expands by a larger factor than if it was ejected at a slower rate~\citep{Kritos:2024upo}. Consequently, when the BH subsystem forms, the escape velocity has dropped by a factor of $\sim10$, and the probability of ejection due to a GW kick is much higher (although the interaction rate also drops). 
Forming a massive BH through BH mergers requires a much higher escape velocity in this scenario, and the initial escape velocity of the cluster has to be larger ($\gtrsim 400\,\rm km\, s^{-1}$) because of the rapid expansion, which lowers $v_{\rm esc}$.
Thus, most of the parameter space in this case does not lead to the production of massive BHs. 

\subsection{Major merger episodes}
\label{sec:Major_merger_episodes}

We keep track of the merger times of galaxies in the merger tree. 
We classify the merger as major or minor based on a decision boundary depending on the mass ratio $Q$ of their stellar masses (defined here as the secondary over the primary galaxy stellar mass).
If the mass ratio is larger than some threshold value $Q_{\rm th}$, we assume that the merger is major, and that if both galaxies contain an NSC, their corresponding NSCs will coalesce. If both NSCs contain massive BHs, these massive BHs will pair up and eventually merge. 
On the other hand, if the asymmetry in the galaxy masses is smaller, we assume that the lighter galaxy dissolves in the primary halo, and the NSC becomes an ultra-compact dwarf galaxy with a central massive BH. 

Following the merger of two galaxies, in case of a major merger, the two NSCs are merged into a single NSC, assuming conservations of mass and energy. 
Hence, if $M_1$, $r_1$ and $M_2$, $r_2$ are the total masses and half-mass radii of the two progenitor NSCs, then the mass $M'$ and half-mass radius $r'$ of the merged NSC will be $M'=M_1+M_2$ and $r'=(M_1^2/r_1 + M_2^2/r_2)^{-1}M'^2$, respectively, as in~\cite{Lupi:2014vza}. 
Moreover, we add an amount of gas given by $f_{\rm g}M'/(1-f_{\rm g})$, where $f_{\rm g}$ is randomly sampled from $[0,f_{\rm g,max}]$, and the choice of $f_{\rm g,max}$ will be discussed later.
In this model, we ignore the effect of any time delays between galaxy mergers, NSC mergers, and massive BH binary mergers. 

Following the coalescence of two massive BHs, we compute the remnant mass, remnant spin, and GW kick $v_{\rm GW}$ imparted to the merger remnant using the fitting formulae of Sec.~V from~\cite{Gerosa:2016sys}, which are based on interpolation between numerical relativity fits and the test particle limit. 
The post-merger remnant BH is retained (ejected) if the magnitude of the GW recoil velocity, $v_{\rm GW}$, is smaller (larger) than the escape velocity of the merged NSC, $v_{\rm esc}$. 
We do not distinguish between massive BHs that receive a superkick (in the thousands of $\rm km\, s^{-1}$) and those that are marginally ejected with $v_{\rm GW}$ barely larger than $v_{\rm esc}$: if $v_{\rm GW}>v_{\rm esc}$, we assume that the ejected massive BH never returns to the center, but rather wanders freely. 

To summarize, massive BHs in our simulations are characterized as being of the following kinds:
\begin{itemize}
\item {\bf Nuclear.} The massive BH lies in the galaxy's center and is surrounded by an NSC.
\item {\bf Off-nuclear/Wandering.} These BHs, in turn, can be either:
  \begin{itemize}
  \item {\bf Satellite}: The massive BH lies in the center of an ultra-compact dwarf hosted in the halo of the primary galaxy following a minor merger episode.
  \item {\bf Ejected}: The massive BH is ejected from the nuclear region following a sufficiently large GW kick.
  \end{itemize}
\end{itemize}

\section{Results}
\label{sec:Results}

In this section, we present the model's predictions for the properties of massive BHs and NSCs in the local Universe (Sec.~\ref{sec:Local_massive_BH_properties} and Sec~\ref{sec:Local_NSC_properties}, respectively). In Sec.~\ref{sec:Transient_events} we discuss transient events, such as massive and stellar-mass BH binary mergers and TDEs. 
We use the merger trees discussed in Sec.~\ref{sec:Merger_trees}. 
Each merger tree terminates at $z=0.25$, and our results are shown at that epoch.
In Table~\ref{tab:fiducial_hyperparameters} we list the set of fiducial hyperparameters used in the simulations.
We run the model for each merger tree once, using a different seed number.

\begin{table}[h]
    \centering
    \caption{Set of fiducial hyperparameter values.}
    \begin{tabular}{c c c c c}
        \hline
        \hline
        $M_{\rm cr}$ & $Q_{\rm th}$ & $\tau_{\rm ge}$ & $f_{\rm g,max}$ & $f_{\star,\rm max}$ \\
        $(M_\odot)$ & & $(\rm Myr)$ & & \\
        \hline
        $10^7$ & $1$:$10$ & $100$ & 1.0 & 1.0 \\
        \hline
    \end{tabular}
    \label{tab:fiducial_hyperparameters}
\end{table}

\subsection{Local massive BH properties}
\label{sec:Local_massive_BH_properties}

In this subsection we discuss the properties of all massive BHs in our $(16\,\rm Mpc)^3$ comoving simulation box at redshift $z=0.25$.
We present, in order:
the $M_{\rm BH}$--$M_\star$ relation (Sec.~\ref{sec:The_MBH_Mstar_relation});
the massive BH spin (Sec.~\ref{sec:Massive_BH_spin}); 
the occupation fraction of massive BHs as a function of galaxy's stellar mass (Sec.~\ref{sec:Occupation_fraction});
and finally, the number densities of IMBHs and SMBHs and their mass function (Sec.~\ref{sec:Number_density}).

\begin{figure}
    \centering
    \includegraphics[width=0.49\textwidth]{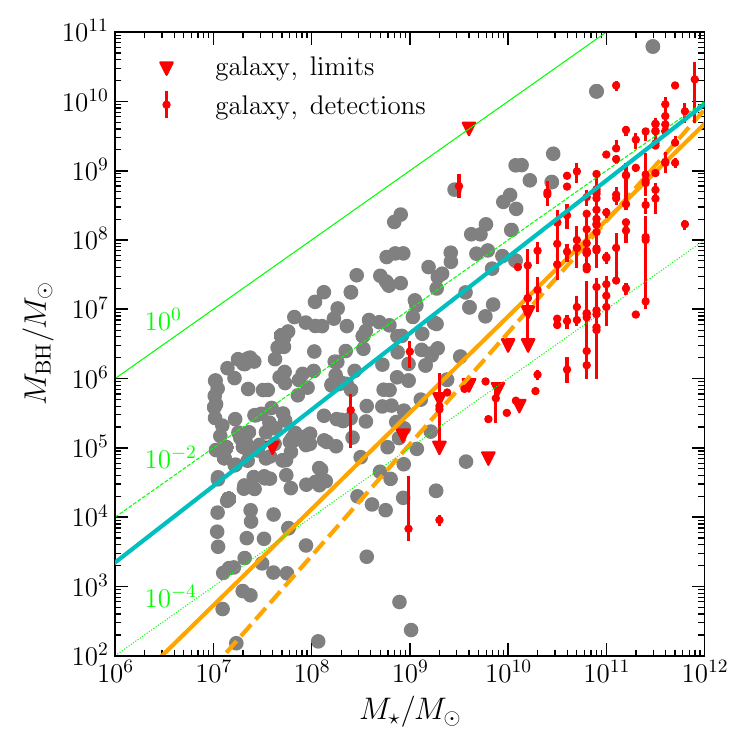}
    \caption{The mass of nuclear massive BHs and the corresponding stellar mass of their host galaxies at redshift $z=0.25$ from the catalog of~\cite{2020ARA&A..58..257G}. The red points with error bars show BH mass estimates in nearby galaxies, while the red triangles are upper limits on BH masses. The cyan solid and orange solid (dashed) lines are power-law fits of the gray and red points including (or excluding) upper limits. For reference, the thin lime lines correspond to different values of $M_{\rm BH}$:$M_{\star}$.}
    \label{fig:MBH-Mstar}
\end{figure}

\subsubsection{The $M_{\rm BH}$--$M_\star$ relation}
\label{sec:The_MBH_Mstar_relation}

In Fig.~\ref{fig:MBH-Mstar} we show all nuclear massive BHs in our simulation box in the $M_{\rm BH}$--$M_{\star}$ plane, where $M_\star$ is the stellar mass of the host galaxy (this is just the result of one realization of NSCs, because $f_\star$ and $f_{\rm g}$ are drawn randomly).
We also include observational data (red points for detections and red triangles for limits) from~\cite{2020ARA&A..58..257G}. Our results do not contain galaxies with $M_\star<10^7\,M_\odot$ because we set the critical stellar mass for NSC formation $M_{\rm cr}=10^7\,M_\odot$ in the merger tree.

If we fit a linear model to our simulation data for $(\log M_{\rm BH}/M_\odot, \log M_\star/10^{10}\,M_\odot)$ using the {\tt linregress} function from {\tt SciPy} and we include only massive BHs with $M_{\rm BH}>100\,M_\odot$ (i.e., we restrict the fit to galaxies which contain a nuclear massive BH), we obtain (note that the logarithm is base 10):
\begin{align}
    \log\left({M_{\rm BH}\over M_\odot}\right) = (7.8\pm0.1) + (1.10\pm0.05)\log\left({M_\star \over 10^{10}\,M_\odot}\right),
    \label{eq:MBH-Mstar_fit}
\end{align}
with a Pearson correlation coefficient of $\simeq0.74$, which implies a strong correlation. The cyan line in Fig.~\ref{fig:MBH-Mstar} corresponds to the mean slope and intercept from this simple model. For comparison, orange lines are similar fits of the observational data from~\cite{2020ARA&A..58..257G} that either include upper limits (solid orange) or consider only estimated masses (dashed orange -- i.e., in this case we exclude systems for which we only have upper limits).

On average, our model predicts larger BH masses relative to the observational data, although there is significant scatter in the data. 
It also predicts a shallower correlation, which roughly follows the $M_{\rm BH}\sim 10^{-2}M_\star$ relation. Thus, our population of massive BHs is heavier in the dwarf galaxy regime $M_\star<10^{9}\,M_\odot$ (bottom-heavy). Moreover, the majority of our galaxy sample are dwarfs with $M_\star<10^9\,M_\odot$. Our $M_{\rm BH}$--$M_\star$ relation has more scatter in the IMBH mass range ($10^2\,M_\odot<M_{\rm BH}<10^6\,M_\odot$), and the $M_{\rm BH}/M_\star$ ratio is in the range $10^{-3}$--$10^{-1}$.
The relation becomes tighter in the SMBH regime ($M_{\rm BH}>10^6\,M_\odot$) and for massive galaxies ($M_\star > 10^{9}\,M_\odot$). For context, the Milky Way has $M_\star\sim10^{11}\,M_\odot$ with a central BH of $\sim10^6\,M_\odot$, and it is therefore an atypical massive galaxy with a relatively small SMBH compared to the average.

\begin{figure}
    \centering
    \includegraphics[width=0.49\textwidth]{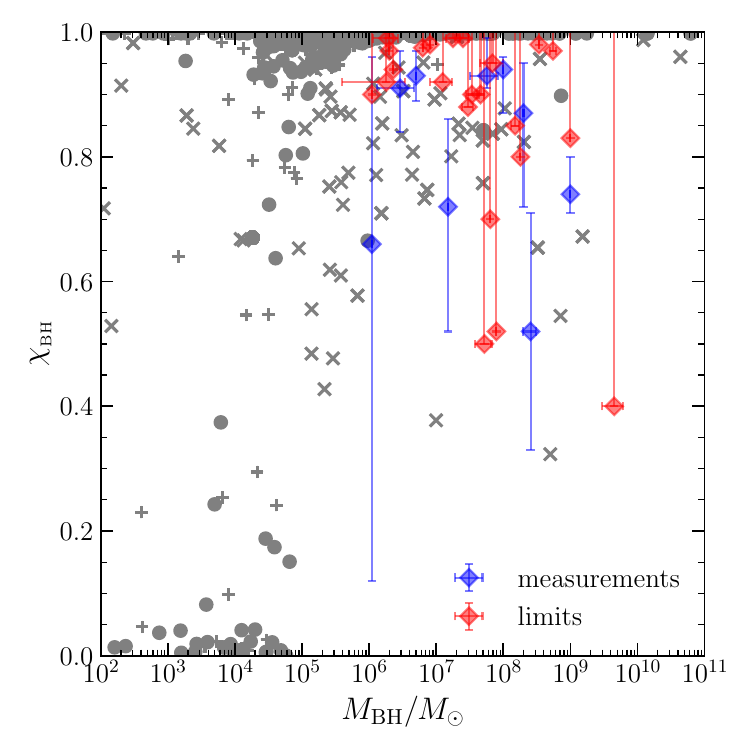}
    \caption{The population of simulated massive BHs at redshift $z=0.25$ in the mass-spin plane. The gray filled circles correspond to nuclear BHs, the ``+'' to satellite BHs, and the ``$\times$'' to ejected massive BHs. We also show measurements with a meaningful upper bound (blue diamonds) and lower limits on spin measurements (red diamonds) made with the X-ray reflection spectroscopy technique. All data are from~\cite{Reynolds:2020jwt}.}
    \label{fig:xBH-MBH}
\end{figure}

\subsubsection{Massive BH spin}
\label{sec:Massive_BH_spin}

In Fig.~\ref{fig:xBH-MBH} we show our simulated massive BH population in the mass-spin plane.
We decompose our dataset into nuclear BHs (gray filled circles), satellite BHs (gray ``+''), and ejected BHs (gray ``$\times$'').
For comparison, we also overplot spin estimates (blue diamonds) and lower limits (red diamonds) made with the X-ray reflection technique, as listed in Table~1 of~\cite{Reynolds:2020jwt}. 
Notice that these published measurements are only in the SMBH regime ($M_{\rm BH}>10^6\,M_\odot$) and they represent a biased population due to selection effects: 
highly spinning SMBHs are easier to measure with the X-ray reflection technique because they are more luminous. 
At the time of writing, no confident spin measurements of IMBH candidates have been published~\citep{2020ARA&A..58..257G}.

We find that nuclear SMBHs ($M_{\rm BH}>10^6\,M_\odot$) are highly spinning ($\chi_{\rm BH}>0.8$). Most of our nuclear SMBH population has $\chi_{\rm BH}\sim1$. This is because SMBHs grow primarily by accreting significant amounts of gas during major galaxy-galaxy mergers. At each gas accretion episode, we model the accretion phase as coherent, and we assume alignment between the BH spin and the angular momentum of the disk.
If a BH coherently accretes an amount of gas at least of the order of its mass, then its spin asymptotes to its maximal value in a short finite time~\citep{Bardeen:1970zz}. For context, the mass increases by one e-fold in about $50\,\rm Myr$, assuming the Eddington accretion rate and a radiative efficiency of 10\%.

The spins of IMBHs are generally not as high. The distribution is bimodal, with one cluster of values around $\chi_{\rm BH}\sim0$ and another one at $\chi_{\rm BH}\sim1$. The spin of IMBHs depends on their formation channel and on their host galaxy merger history.
If most of the IMBH's mass is assembled through repeated BH mergers or consumption of stars, the spin asymptotes to zero~\citep{Hughes:2002ei}. This results from the asymmetry of the innermost stable circular radius for prograde vs. retrograde orbits and the assumption of isotropy. On the other hand, coherent gas accretion events spin up the BH to unity as long as the BHs accrete an amount of gas $\sim M_{\rm BH}$~\citep{Bardeen:1970zz,Berti:2008af,Dotti:2012qw,Dubois:2014yya,Beckmann:2024ezt,Kritos:2024kpn}. 

\begin{figure}
    \centering
    \includegraphics[width=0.49\textwidth]{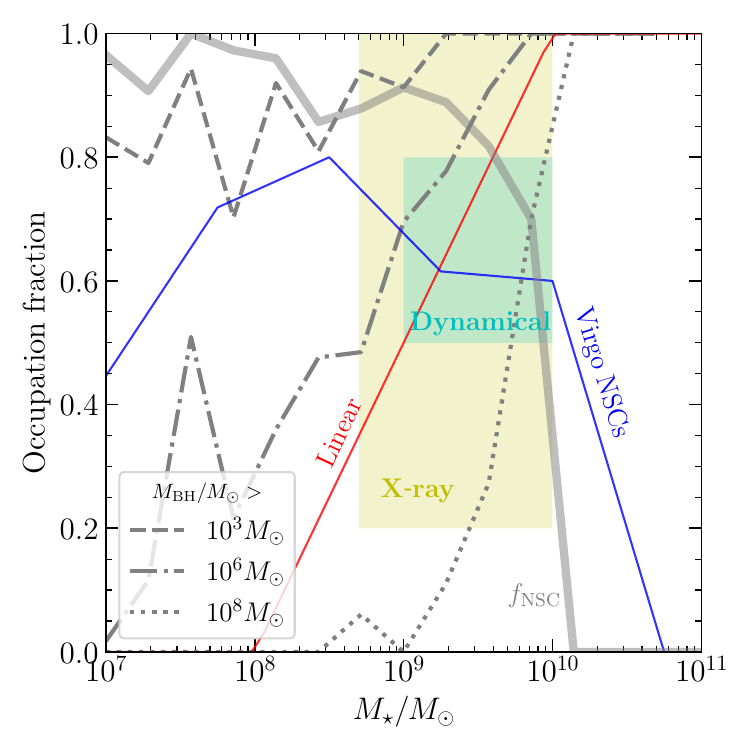}
    \caption{Occupation fraction of massive BHs with mass $>10^3\,M_\odot$ (gray dashed), $>10^6\,M_\odot$ (gray dash-dotted), and $>10^8\,M_\odot$ (gray dotted), as well as the NSC occupation fraction ($f_{\rm NSC}$, thick gray solid). The linear model of~\cite{2020ARA&A..58..257G} for the $M_{\rm BH}>10^5\,M_\odot$ occupation fraction and the NSC occupation fraction in the Virgo cluster from~\cite{2019ApJ...878...18S} are shown with red and blue lines, respectively. The cyan- and yellow-filled regions correspond to massive BH occupation fraction limits from the dynamical and X-ray surveys in~\cite{2019ApJ...872..104N} and~\cite{Miller:2014vta}, respectively.}
    \label{fig:focc-Mstar}
\end{figure}

\subsubsection{Occupation fraction}
\label{sec:Occupation_fraction}

In Fig.~\ref{fig:focc-Mstar} we show the occupation fraction of massive nuclear BHs with mass $>10^3\,M_\odot$ (gray-dashed), $>10^6\,M_\odot$ (gray-dash-dotted), and $>10^8\,M_\odot$ (gray-dotted), as well as the occupation fraction $f_{\rm NSC}$ of NSCs with half-mass radii $<50\,\rm pc$ (thick gray solid line).
We assume any NSC with a half-mass radius that exceeds $50\,\rm pc$ to have dissolved in the host galaxy and contributed to the bulge component.
For comparison, we show limits on the BH occupation fraction in low-mass galaxies with masses $10^9$--$10^{10}\,M_\odot$ from dynamical~\citep{2019ApJ...872..104N} and X-ray~\citep{Miller:2014vta} surveys [see also Fig.~5 in~\cite{2020ARA&A..58..257G}].
For reference, the blue line shows the nucleation fraction of galaxies in the Virgo cluster from the catalog of~\cite{2019ApJ...878...18S}, and the red line is the linear model for the occupation fraction of BHs with $M_{\rm BH}>10^5\,M_\odot$ from~\cite{2020ARA&A..58..257G}.

Our model predicts that $\sim80\%$ of the dwarf galaxies with stellar mass $\lesssim10^9\,M_\odot$ contain IMBHs with mass $>10^3\,M_\odot$.
A fraction $<20\%$ of these dwarf galaxies contain SMBHs ($M_{\rm BH}>10^6\,M_\odot$).
Massive BHs with $M_{\rm BH}>10^8\,M_\odot$ are hosted by galaxies with stellar mass $>10^9\,M_\odot$.
The occupation fraction converges to unity in massive galaxies with stellar mass $\gtrsim10^{10}\,M_\odot$.
For comparison, the value of the occupation fraction for BHs with $M_{\rm BH}>10^5\,M_\odot$ predicted by other studies is in the range $0.1$--$0.7$ ($0.1$--$1.0$) for $10^8\,M_\odot$ ($10^9\,M_\odot$) stellar-mass galaxies [see Table~1 of~\cite{2020ARA&A..58..257G}].
Our predictions tend to be on the higher end of these ranges.

The NSC occupation fraction approaches unity for dwarfs with stellar mass below $10^8\,M_\odot$ and drops to zero in massive galaxies above $10^{10}\,M_\odot$ due to SMBH feedback, which contributes to the dissolution of these NSCs through adiabatic expansion during gas removal. In contrast, as shown by the right panel of Fig.~3 based on observational data in~\cite{2020A&ARv..28....4N}, the occupation declines for dwarfs. However, it should be noted that the results on the NSC occupation fraction presented in~\cite{2020A&ARv..28....4N} are lower limits, due to the difficulty of identifying NSCs.
Our predictions for high occupation fractions are impacted by our choice to form an NSC in every galaxy with stellar mass larger than $10^7\,M_\odot$.

\begin{figure}
    \centering
    \includegraphics[width=0.49\textwidth]{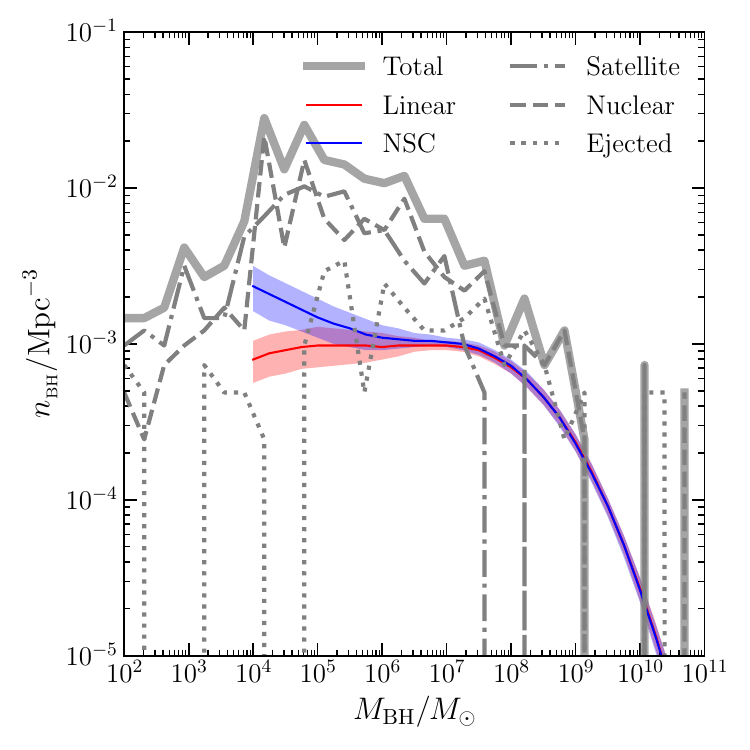}
    \caption{Local massive BH mass function, decomposed into satellite (dash-dot), nuclear (dashed), and ejected (dotted) BH components. The thick solid line corresponds to the total number density. The lower limit estimates ``Linear'' and ``NSC'' from~\cite{2020ARA&A..58..257G} for the nuclear contribution of the mass function are shown in red and blue, respectively.}
    \label{fig:nBH-MBH}
\end{figure}

\subsubsection{Number density}
\label{sec:Number_density}

In Fig.~\ref{fig:nBH-MBH} we show the local massive BH mass function at $z=0.25$.
We decompose the total number density into satellite, nuclear, and ejected BH components. For comparison, we show the lower limit estimates of the mass function in Fig.~6 from~\cite{2020ARA&A..58..257G}, which assumes two models for the occupation fraction: the ``Linear'' and ``NSC'' model.

In total, the simulation box contains 522 (201) IMBHs (SMBHs) in the $(16\,\rm Mpc)^3$ comoving volume at $z=0.25$, corresponding to a local number density of $0.13\,\rm Mpc^{-3}$ ($4.9\times10^{-2}\,\rm Mpc^{-3}$).
Moreover, the number density of ejected IMBHs (SMBHs) is $1.1\times10^{-2}\,\rm Mpc^{-3}$ ($1.4\times10^{-2}\,\rm Mpc^{-3}$).
We find roughly equal contributions ($\approx6.4\times10^{-2}\,\rm Mpc^{-3}$) of nuclear and satellite IMBHs. For comparison, according to other models discussed in Sec.~2.4 of~\cite{2020ARA&A..58..257G}, the number density of IMBHs in nuclei is in the range $0.02$--$0.25\,\rm Mpc^{-3}$. This is consistent with our result for the density of nuclear IMBHs. On the other hand, the number density of wandering IMBHs -- what we term ``off-nuclear,'' which combines the satellite and ejected components -- is predicted to be $>0.3\,\rm Mpc^{-3}$.
Consequently, our model predicts a factor of at least four lower number densities of off-nuclear IMBHs than other studies.

The BH mass function turns over at $M_{\rm BH}\sim10^{4}\,M_\odot$, and indicates that our NSC scenario produces a larger abundance of heavy seeds with masses $\sim10^5\,M_\odot$. 
Most of these IMBHs are assembled in dwarf galaxies that do not have a rich merger history. The IMBHs stay in the center of their hosts until the present era.
There is an underdensity of light seeds in the range $10^2$--$10^4\,M_\odot$, while the majority of BHs in the Universe are stellar-mass BHs, remnants of massive stars that did not grow.
We summarize the number densities discussed in this section in Table~\ref{tab:Massive_BH_number_densities}.
Notice that our IMBH densities have been derived assuming that they assemble solely in NSCs. In contrast, several IMBHs might have been formed in off-nuclear globular star clusters or other environments. Therefore, our IMBH number density estimates represent lower limits.
Finally, given the small volume of \nh there are no massive galaxies; therefore, the number density is a lower limit at the high SMBH mass end.

\begin{table}
    \centering
    \caption{Predicted number densities of massive BHs (IMBHs and SMBHs) at $z=0.25$.}
    \begin{tabular}{c c c}
        \hline
        \hline
         Massive BH & $n_{\rm IMBH}$ & $n_{\rm SMBH}$ \\
         type & ($\rm Mpc^{-3}$) & ($\rm Mpc^{-3}$) \\ \hline
        Nuclear & 0.064 & 0.032 \\
        Satellite & 0.064 & 0.017 \\ 
        Ejected & 0.011 & 0.014 \\ \hline
        Total & 0.130 & 0.049 \\ \hline
    \end{tabular}
    \label{tab:Massive_BH_number_densities}
\end{table}

\subsection{Local NSC properties}
\label{sec:Local_NSC_properties}

In this subsection we discuss the present-day properties of NSCs within our simulation box at redshift $z=0.25$. We consider only NSCs with a final half-mass radius smaller than $50\,\rm pc$, as we assume that sparser NSCs have dissolved and contribute to the bulge component of their host. In total, the $(16\,\rm Mpc)^3$ comoving box contains 1133 NSCs with $r_{\rm h}<50\,\rm pc$, corresponding to a local NSC number density of roughly $0.28\,\rm Mpc^{-3}$. This is comparable to (but a factor of two larger than) the predicted total local number density of massive BHs ($M_{\rm BH}>100\,M_\odot$). In the following, we present the $M_{\rm BH}$--$M_{\rm NSC}$ relation (Sec.~\ref{sec:The_MBH_Mnsc_relation}) as well as the $r_{\rm h}$--$M_{\rm NSC}$ and $M_{\rm NSC}$--$M_\star$ relations (Sec.~\ref{sec:The_rh-Mnsc_relation}) predicted by the model.

\begin{figure}
    \centering
    \includegraphics[width=0.49\textwidth]{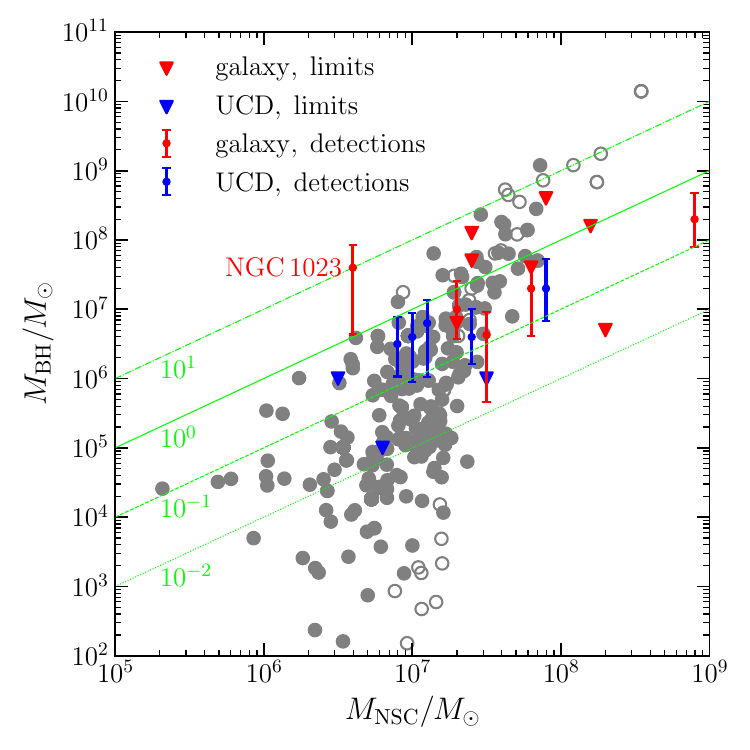}
    \caption{The BH mass vs. the total mass of the host NSC for the population of massive BHs at redshift $z=0.25$. The hollow circles correspond to NSCs whose half-mass radius has expanded past $50\,\rm pc$. We also plot observational data from~\cite{2020ARA&A..58..257G}, including claimed detections (points with error bars) and upper mass limits (triangles) in galactic centers (red) and ultra-compact dwarfs (UCDs, blue). For reference, the thin lime lines show different values of $M_{\rm BH}/M_{\rm NSC}$.}
    \label{fig:MBH-Mnsc}
\end{figure}

\begin{figure*}[t]
    \centering
    \includegraphics[width=0.49\linewidth]{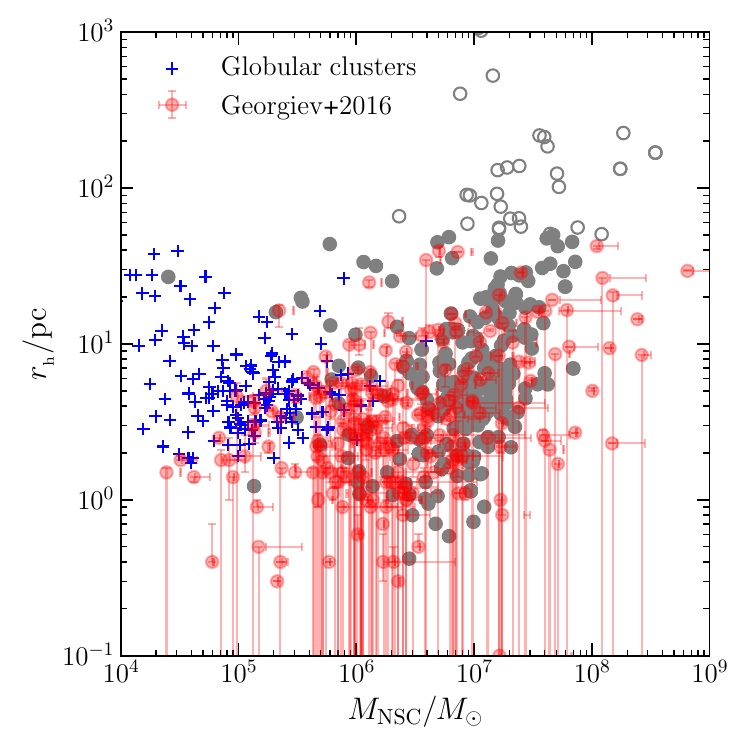}
    \includegraphics[width=0.49\linewidth]{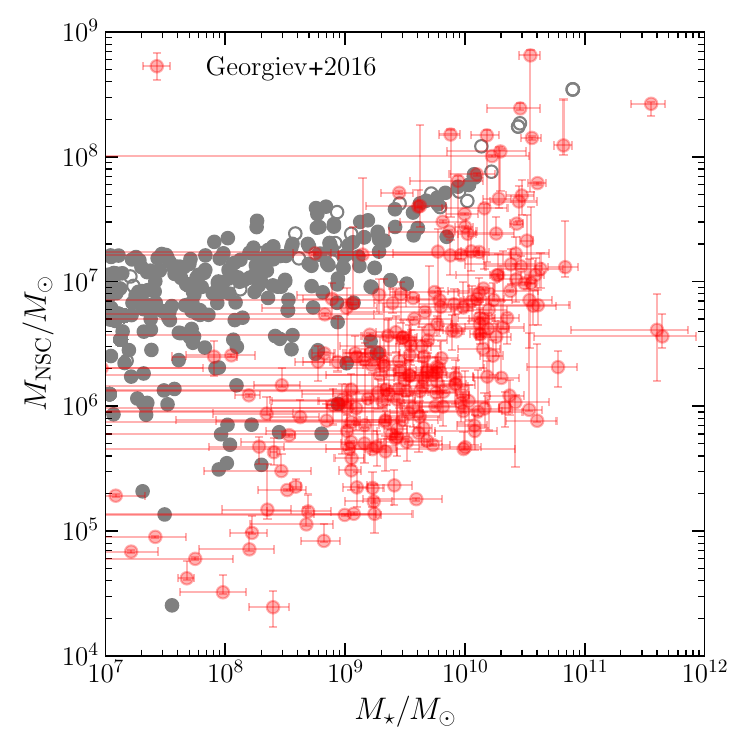}
    \caption{Left: Present-day structural properties of NSCs (gray points) in the half-mass radius ($r_{\rm h}$) versus NSC mass ($M_{\rm NSC}$) plane. Also shown are Milky-Way globular clusters (blue crosses) and the NSC catalog of~\cite{2016MNRAS.457.2122G} (red points with error bars). Right: same as the left panel, but in the galaxy stellar mass ($M_\star$) vs. NSC mass ($M_{\rm NSC}$) plane.}
    \label{fig:rh-Mnsc}
\end{figure*}

\subsubsection{The $M_{\rm BH}$--$M_{\rm NSC}$ relation}
\label{sec:The_MBH_Mnsc_relation}

In Fig.~\ref{fig:MBH-Mnsc}, we show the local $M_{\rm BH}$--$M_{\rm NSC}$ correlation. For comparison, we include BH and NSC data from~\cite{2020ARA&A..58..257G} corresponding to either claimed detections or upper limits in the centers of galaxies and ultra-compact dwarfs.

Despite the large scatter in the data and in the simulation results, we observe a correlation between the BH mass and the NSC mass, because heavier clusters harbor heavier BHs.
In particular, BHs with mass $>10^6\,M_\odot$ and $>10^8\,M_\odot$ are only hosted in NSCs with mass $\gtrsim10^6\,M_\odot$ and $\gtrsim10^7\,M_\odot$, respectively.
The scatter increases in NSCs with mass $\lesssim10^7\,M_\odot$, and the results span a wide range of values of $M_{\rm BH}/M_{\rm NSC}\in [10^{-4},10]$.
The observational data also suggest a positive correlation between BH and NSC mass, albeit with a shallower slope. However, due to selection biases, the comparison is only qualitative.

The heaviest NSCs in our catalog ($M_{\rm NSC}\gtrsim10^8\,M_\odot$) contain overmassive SMBHs, in the sense that the BH mass is larger than the NSC mass, in some cases up to a factor of $\sim10$. Most of these NSCs have half-mass radii larger than $50\,M_\odot$, which we define as the threshold for NSC detectability.
A special case is NGC 1023, which may contain an SMBH whose mass exceeds the host NSC mass by a factor of $\sim10$ or more. However the error bar on $M_{\rm BH}$ is large, and the $M_{\rm BH}/M_{\rm NSC}$ ratio could be as high as $\sim1$. 

\subsubsection{The $r_{\rm h}$--$M_{\rm NSC}$ and $M_{\rm NSC}$--$M_\star$ relations}
\label{sec:The_rh-Mnsc_relation}

In Fig.~\ref{fig:rh-Mnsc}, we show the structural properties of our NSC population in the local Universe. We mark with hollow gray circles all NSCs with a half-mass radius that has expanded past $50\,\rm pc$. We further compare our results against the NSC catalog of~\cite{2016MNRAS.457.2122G} and the population of Milky-Way globular clusters (in the left panel).
Our NSC population is heavier than the sample from~\cite{2016MNRAS.457.2122G}. Moreover, NSCs are heavier than globular clusters by at least an order of magnitude (on average) but have similar half-mass radii, as confirmed by observations~\citep{2020A&ARv..28....4N}.

\subsection{Transient events}
\label{sec:Transient_events}

In this subsection we compute the merger rates of massive BH mergers (Sec.~\ref{sec:Massive_BH_binary_merger_rate}),
stellar-mass BH-BH captures in NSCs (Sec.~\ref{sec:Stellar_mass_BH-BH_capture_rate}), extreme-mass ratio inspirals (Sec.~\ref{sec:Extreme-mass_ratio_inspirals}), and TDEs (Sec.~\ref{sec:Tidal_disruption_event_rate}).

\begin{figure}
    \centering
    \includegraphics[width=0.49\textwidth]{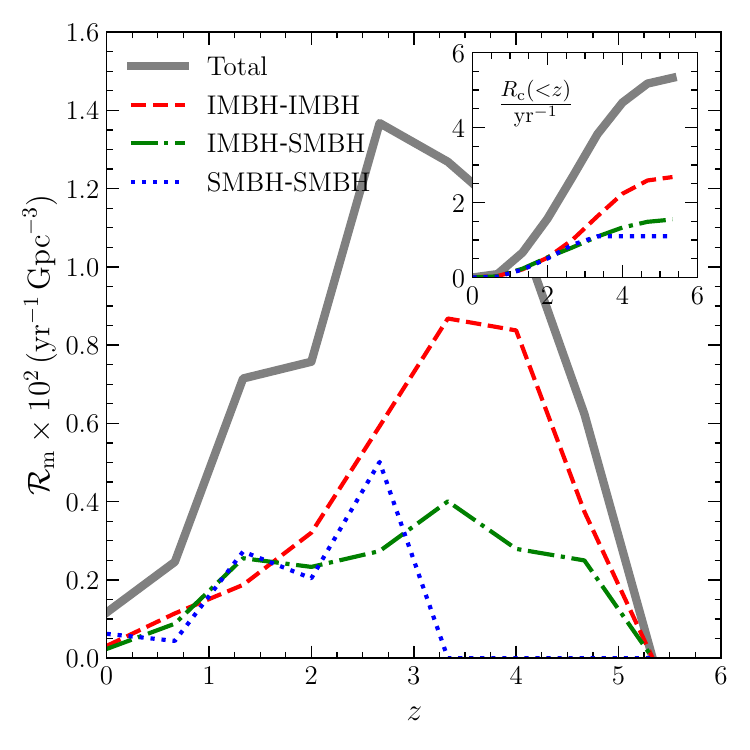}
    \caption{Source-frame merger rate density of massive BHs as a function of redshift (gray). We decompose this rate into IMBH-IMBH (red-dashed), IMBH-SMBH (green-dash-dotted), and SMBH-SMBH (blue-dotted) merger contributions. The inset shows the cumulative observed merger rate out to redshift $z$.}
    \label{fig:Rm_Rc_MBHMBH-z}
\end{figure}

\subsubsection{Massive BH binary merger rate}
\label{sec:Massive_BH_binary_merger_rate}

Massive BH binaries form when galaxies merge with a minor-to-major galaxy mass ratio that does not exceed a certain threshold, for which we take the fiducial value $Q_{\rm th}=0.1$. Thus, the massive BH binary merger rate in our model follows the major galaxy-galaxy merger rate. 

In Fig.~\ref{fig:Rm_Rc_MBHMBH-z} we show the source-frame massive BH binary merger rate as a function of redshift, ${\cal R}_{\rm m}(z)$, as well as its decomposition into the three subchannels: IMBH-IMBH, IMBH-SMBH, and SMBH-SMBH.  The inset shows the observer-frame cumulative merger rate up to redshift $z$, computed using Eq.~(17) from~\cite{Rodriguez:2016kxx}:
\begin{align}
    R_{\rm c}(<z)=\int_0^{z} {\cal R}_{\rm m}(z'){dV_{\rm com}(z')\over dz'}{dz'\over1+z'},
    \label{eq:Rc}
\end{align}
where $dV_{\rm com}(z)$ is the all-sky differential comoving volume element at redshift $z$.
We find that the total merger rate attains its maximum at $z\simeq3$, with a peak value $\simeq1.4\times10^{-2}\,\rm yr^{-1}\, Gpc^{-3}$. The corresponding values for each subchannel are: $\simeq0.8\times10^{-2}\,\rm yr^{-1}\, Gpc^{-3}$ at $z\simeq4.0$ for IMBH-IMBH, $\simeq0.4\times10^{-2}\,\rm yr^{-1}\, Gpc^{-3}$ at $z\simeq3.5$ for IMBH-SMBH, and $\simeq0.5\times10^{-2}\,\rm yr^{-1}\, Gpc^{-3}$ at $z\simeq2.5$ for SMBH-SMBH mergers. Hence, the total source-frame mass decreases with redshift, as expected from a hierarchically merging massive BH population as larger structures assemble with time. In the local Universe, the SMBH-SMBH merger rate is predicted to be of the order of a few\,$\times10^{-2}\,\rm yr^{-1}\, Gpc^{-3}$. The observer-frame merger rate out to $z=5$ is $\simeq2.7\,\rm yr^{-1}$ for IMBH-IMBH, $\simeq1.5\,\rm yr^{-1}$ for IMBH-SMBH, and $\simeq1.1\,\rm yr^{-1}$ for SMBH-SMBH mergers, adding up to a total of $\simeq5.3\,\rm yr^{-1}$.

\begin{figure}
    \centering
    \includegraphics[width=\linewidth]{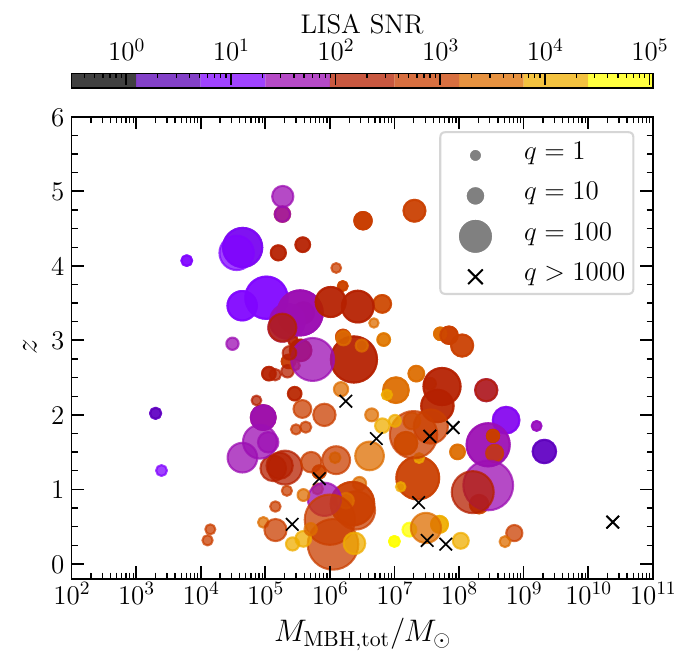}
    \caption{Massive BH binary mergers in our generated catalog in the plane of merger redshift $z$ and total source-frame binary mass $M_{\rm MBH, tot}$. The color shows the signal-to-noise ratio (SNR) of each merger event in the LISA detector. The point size represents $\sqrt{q}$, where $q=m_1/m_2>1$ is the binary mass ratio. Benchmark point sizes for $q=[1,10,100]$ are given in the legend, and crosses correspond to merger events for which $q>1000$, such that our waveform models cannot be trusted for the SNR calculation.}
    \label{fig:M_z_rho_q}
\end{figure}

In Fig.~\ref{fig:M_z_rho_q} we show the properties of our massive BH binary merger catalog. These GW sources are observational targets for space-borne GW observatories such as LISA~\citep{LISA:2017pwj}. In total, the catalog contains 192 events. For each event with a primary-to-secondary mass ratio of $q<1000$, we compute the signal-to-noise (SNR) ratio in the LISA detector using the \texttt{lisabeta} software~\citep{Marsat:2020rtl} with the {\tt IMRPhenomXHM} waveform~\citep{Garcia-Quiros:2020qpx}, assuming aligned spins for simplicity. For each event, we select random values for the source location, inclination, and polarization angles, although these values do not significantly alter the detectability prospects. We find that $92\%$ of all events have a signal-to-noise ratio $>10$ out to redshift $z=6$. The loudest events have total source-frame masses in the range $10^6$--$10^7\,M_\odot$, roughly equal mass components ($q\simeq 1$), and $z\lesssim2$.

\subsubsection{Stellar mass BH-BH capture rate}
\label{sec:Stellar_mass_BH-BH_capture_rate}

Interactions among stellar-mass BHs around massive BHs in the cores of NSCs lead to the formation of BH binaries that merge. Here, we compute the rate of BH-BH captures through the emission of GWs in NSCs. This rate corresponds to the merger rate, because captured pairs merge promptly, on a time scale smaller than the disruption time scale, as long as the density is $\lesssim10^{12}\,{\rm pc}^{-3}\,(v_{\rm rel}/100\,\rm km\,s^{-1})^6$~\citep{OLeary:2008myb}. The two-body GW capture cross-section, $\Sigma_{\rm cap}$, is given by Eq.~(4) from~\cite{Mouri:2002mc}, where we set $m_1=m_2=10\,M_\odot$ and $v_{\rm rel}=\sqrt{2}v_{\rm BH}$ (where $v_{\rm BH}$ is the three-dimensional velocity dispersion in the BH subsystem). Assuming for simplicity uniform conditions within the half-mass radius of the BH subsystem, $r_{\rm BH}$, which is related to $v_{\rm BH}$ through the virial theorem, we compute the capture rate per NSC as
\begin{align}
    \Gamma_{\rm cap} &\approx {N_{\rm BH}^2\over4r_{\rm BH}^3}\Sigma_{\rm cap}v_{\rm rel}\nonumber\\&\simeq 2.6\times10^{-5}\,{\rm yr}^{-1}\,{1000\over N_{\rm BH}}\left({v_{\rm BH}\over100\,\rm km\, s^{-1}}\right)^{31/7}.
\end{align}
Here $N_{\rm BH}$ is the number of BHs in the BH subsystem.
As the NSC expands, the velocity dispersion drops, and the capture rate declines with time.
To compute the total source-frame capture rate density at redshift $z$, we add up the contributions from each NSC within the simulated comoving volume and divide by that volume:
\begin{align}
    {\cal R}_{\rm cap}(z) = {1\over(16\,\rm Mpc)^3}\sum_{i} \Gamma_{{\rm cap},i}(z),
    \label{eq:R_cap}
\end{align}
where the index $i$ runs over all NSCs for which the time-integrated capture rate does not exceed $N_{\rm BH}/2$. 

\begin{figure}
    \centering
    \includegraphics[width=\linewidth]{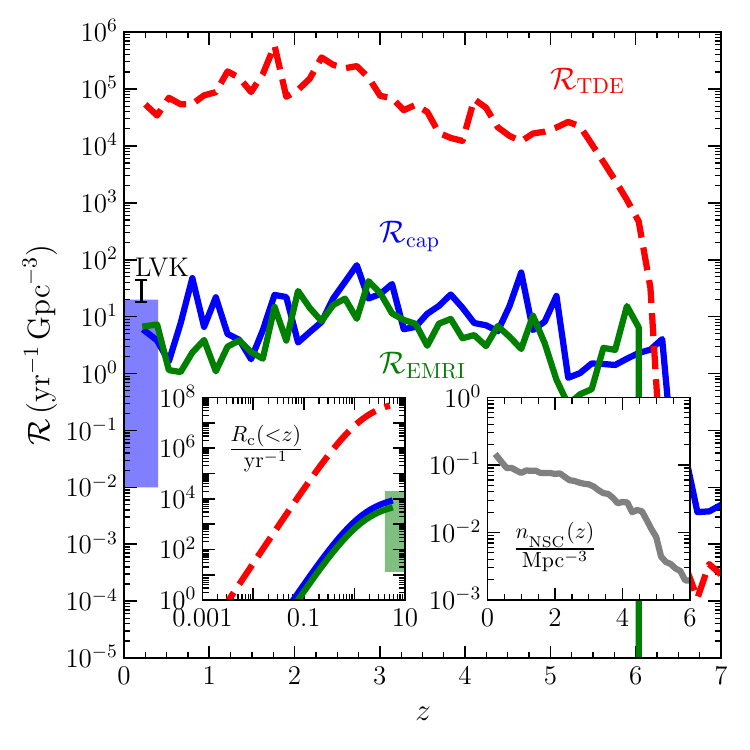}
    \caption{Source-frame rate density of stellar-mass BH-BH captures (solid blue), extreme-mass ratio inspirals (solid green), and TDEs (dashed red) in NSCs as a function of redshift. Their corresponding observer-frame cumulative numbers [$R_{\rm c}(<z)$] and the number density of NSCs [$n_{\rm NSC}(z)$] are shown in the left and right inset plots, respectively. The blue band represents the range of model predictions for stellar-mass BH mergers in NSCs, and the black error bar corresponds to the BH-BH merger rate inferred by the LIGO-Virgo-KAGRA (LVK) GW network. Finally, the green band in the left inset shows the rate predictions for extreme-mass ratio inspirals from~\cite{Babak:2017tow}.}
    \label{fig:Rm_Rc_BHBH_TDE_EMRI-z}
\end{figure}

In Fig.~\ref{fig:Rm_Rc_BHBH_TDE_EMRI-z} we plot the source-frame rate density of BH-BH captures as a function of redshift (solid blue), as well as the observed intrinsic cumulative merger rate (left inset) and number density of contributing NSC, $n_{\rm NSC}(z)$ (right inset). Due to finite number statistics, the estimated rate density shows fluctuations; nevertheless, on average, the source frame volumetric capture rate is $\approx10\,\rm yr^{-1}\, Gpc^{-3}$ out to redshift $z\simeq5$. The rate decreases at $z>5$ due to a declining number density of NSCs. Predictions of the local BH-BH merger rate from NSCs in other studies ranges from $\sim10^{-2}$ up to a few tens of events $\rm yr^{-1}\, Gpc^{-3}$ [see Fig.~3 in~\cite{Mandel:2021smh}, and the blue vertical band in Fig.~\ref{fig:Rm_Rc_BHBH_TDE_EMRI-z}] while the inferred merger rate from GW observations during O3 is in the range $18$--$44\,\rm yr^{-1}\, Gpc^{-3}$~\citep[][cf. the black error bar ``LVK'' in Fig.~\ref{fig:Rm_Rc_BHBH_TDE_EMRI-z}]{KAGRA:2021duu}. We compute the cumulative merger rate $R_{\rm c}$ from Eq.~\eqref{eq:Rc} assuming the average source-frame merger rate density. Within $z=5$ , we find the cumulative observed intrinsic rate of BH-BH captures to be $\simeq6000\, \rm yr^{-1}$.
For comparison, it is expected that several hundred thousand BH-BH mergers per year
would occur from the isolated binary evolution channel~\citep[see e.g.][]{Baibhav:2019gxm}. Therefore, BH-BH captures in NSCs represent a fraction of $\sim1\%$ of the total stellar-mass BH binary merger rate. However, merging binaries that formed from captures tend to have eccentricities $\sim10^{-1}$ at a (detector-frame) GW frequency $10\,\rm Hz$, with significant support for eccentricities above $0.1$. Therefore these binaries have different orbital properties than binaries merging in the field~\citep{OLeary:2008myb,Zevin:2021rtf,Fumagalli:2024gko}. These stellar-mass BH-BH captures may be detectable by ground-based GW observatories such as the LIGO-Virgo-KAGRA network, and by future observatories such as the Einstein Telescope and Cosmic Explorer.

\subsubsection{Extreme-mass ratio inspirals}
\label{sec:Extreme-mass_ratio_inspirals}

As massive BHs grow in NSCs through repeated mergers with $10\,M_\odot$ stellar-mass BHs, a population of extreme-mass ratio inspirals contributes to the GW background in the mHz band. Such inspirals are potential LISA sources.
We estimate the total rate density of extreme-mass ratio inspirals with primaries of mass $>10^4\,M_\odot$ (see the green line in Fig.~\ref{fig:Rm_Rc_BHBH_TDE_EMRI-z}).

We find an intrinsic merger rate density comparable to the capture rate between stellar-mass BHs, and a cumulative rate of $\sim4000$ inspirals per year.
For comparison, we show the range of predictions in Table~I from~\cite{Babak:2017tow} (green band in the left inset of Fig.~\ref{fig:Rm_Rc_BHBH_TDE_EMRI-z}), where extreme-mass ratio inspiral rates lie in the range $13$--$20000\,\rm yr^{-1}$, depending on model assumptions.

\subsubsection{Tidal disruption event rate}
\label{sec:Tidal_disruption_event_rate}

Stars approaching BHs within the tidal radius $r_t$ may be tidally disrupted.
The condition for generating an observable electromagnetic counterpart is that $r_s<r_t<r_\star$, where $r_s$ is the BH Schwarzschild radius and $r_\star$ is the star's physical radius, which constrains the BH mass to be in the range $10^5\,M_\odot\lesssim M_{\rm BH}\lesssim10^8\,M_\odot$~\citep{Gezari:2021bmb}.
If these conditions are met, the TDE cross-section in the gravitational focusing regime is given by
\begin{align}
    \Sigma_{\rm TDE}\simeq \pi r_s r_t \left(c\over v_\star\right)^2,
\end{align}
where $v_\star$ is the three-dimensional stellar velocity dispersion.
Using the half-mass number density of stars, the TDE rate per NSC in the full loss cone is computed as~\citep{Pfister:2020quh}
\begin{align}
    \Gamma_{\rm TDE} &\approx {N_\star\over4r_{\rm h}^3}\Sigma_{\rm TDE}v_\star \nonumber\\&\simeq 3\times10^{-4}\,{\rm yr}^{-1}\,\nonumber\\&\times \left({10^7\over N_\star}\right)^2 \left({M_{\rm BH}\over10^6\,M_\odot}\right)^{4/3} \left({v_\star\over100\,\rm km\, s^{-1}}\right)^{5}.
\end{align}
Again, the half-mass radius $r_{\rm h}$ is related to $v_\star$ through the virial theorem. The source-frame volumetric TDE rate density ${\cal R}_{\rm TDE}(z)$ and the observer-frame cumulative rate are computed as in Eq.~\eqref{eq:R_cap} and Eq.~\eqref{eq:Rc}.
The TDE rate density is in the range $10^4$--$10^5\,\rm yr^{-1}\, Gpc^{-3}$ out to $z=5$, about 3--4 orders of magnitude above the BH-BH capture rate density.
Based on the lower-left inset of Fig.~\ref{fig:Rm_Rc_BHBH_TDE_EMRI-z}, we expect thousands of TDEs and a single BH-BH capture in an NSC at a luminosity distance of $500\,\rm Mpc$ ($z\simeq0.1$).

\section{Caveats}
\label{sec:Caveats}

In this section, we list the main limitations of our study (Sec.~\ref{sec:Limitations}) and we qualitatively discuss the dependence of our results on the choice of our model hyperparameters (Sec.~\ref{sec:Dependence_of_hyperparameters}).

\subsection{Limitations}
\label{sec:Limitations}

Below, we list how the assumptions made in our model may affect our results.

The NSC initial conditions presented in Sec.~\ref{sec:NSC_initial_conditions} were arbitrarily chosen. While a few strongly magnified star clusters have been identified with the JWST~\citep{2022ApJ...937L..35M,2023ApJ...943....2W,2023ApJ...945...53V,2024Natur.632..513A}, there is no consensus on the initial conditions for high-redshift NSCs. Compact NSCs on sub-pc scales could theoretically form in the high-redshift Universe ($z>10$)~\citep{2010MNRAS.409.1057D,2016ApJ...823...52K,2024arXiv241100670M}. This assumption is essential for assembling SMBH seeds in the NSC paradigm (cf.~Fig.~\ref{fig:MBH-xBH-t}).

We consider a single burst of NSC formation when the stellar mass of each galaxy reaches the critical value of $M_{\rm cr}=10^7\,M_\odot$. This assumption may lead to an overabundance of NSCs in the local Universe if not every galaxy forms an NSC in its core. While we consider the in-situ formation of NSCs in protogalaxies, we note that globular clusters that sink to the center due to dynamical friction and form near the nuclear region of the host galaxy may contribute to the mass assembly of the NSC [see, e.g., \cite{Gnedin:2013cda}]. In our model, NSCs lose mass as they evolve, and gain mass through NSC mergers and through gas inflows during major mergers. We simplify our analysis by neglecting the contribution from this ex-situ channel of inspiralling globular clusters. Furthermore, NSCs have been observed with multiple stellar populations~\citep{2020A&ARv..28....4N}, indicating complex star formation histories rather than a monolithic formation at some high redshift. 
The larger number of NSC mergers also contributes to the presence of overmassive BHs in our model, because BHs grow by mergers and gas accretion.

The off-nuclear number density we estimated in Sec.~\ref{sec:Number_density} is likely a lower bound, because some off-nuclear star clusters could form IMBHs in their centers. If some of these IMBHs are ejected from their cores [as suggested by some simulations~\citep{Holley-Bockelmann:2007hmm,Prieto:2022uot}, especially for IMBHs with low-mass $M_{\rm BH}\lesssim1000\,M_\odot$], then our estimated number density of ejected massive BHs is also a lower bound.

We have not included the time delay in massive BH binary coalescence following major galaxy mergers (but see~\cite{2024arXiv241007856L} for subgrid modeling). This time may be decomposed into three contributions: the dynamical friction time scale ($\tau_{\rm df}$, from galaxy merger to formation of the tight massive BH binary); the binary hardening time scale ($\tau_{\rm har}$, from binary formation until its evolution becomes dominated by GW emission); and the GW merger time scale ($\tau_{\rm gw}$), so that $\tau_{\rm delay}=\tau_{\rm df}+\tau_{\rm har}+\tau_{\rm gw}$ [see, e.g., Eq.~(6) from~\cite{Langen:2024ygz}]. We assume instantaneous binary formation in the post-merger galaxy by ignoring $\tau_{\rm df}$. Furthermore, we do not consider cases where a third massive BH is brought into the vicinity of the evolving massive BH pair following a third subsequent galaxy-galaxy merger episode. Assuming that the {\it final parsec problem} is not, in fact, a problem~\citep{Khan:2013wbx}, we can estimate the time delay from binary formation until binary merger as [see Eq.~(30) and~(31) from~\cite{Quinlan:1996vp}]
\begin{align}
  \tau_{\rm har}+\tau_{\rm gw}&\sim 44\,{\rm Myr}\, \left({\sigma_\star \over 100\,\rm km\, s^{-1}}\right)^{4/5} \left({H\over16}\right)^{-4/5}\nonumber\\&\times \left({\rho_\star\over 10^5M_\odot\, \rm pc^{-3}}\right)^{-4/5} \left({M_{\rm BH}\over 10^6M_\odot}\right)^{-3/5}.
\end{align}
This time scale can thus be much smaller than the time scale between successive galaxy mergers, and it can be neglected.

Our simulations are built on merger tree outputs from the \nh results. Hence, no feedback effect from the massive BH or the NSC on the merger tree has been implemented.

Our population of SMBHs is highly spinning. In principle, the presence of strong magnetic fields in the vicinity of SMBHs may efficiently extract angular momentum from the BH and slow it down through the Blandford-Znajek process~\citep{Blandford:1977ds}. We have not simulated this process. The highly spinning SMBHs in our simulations are consistent with our assumed coherent gas accretion episodes. On the other hand, if the accretion process is chaotic, then the spin asymptotes to zero~\citep{King:2008au}.

We have limited the accretion rate to the Eddington limit. A subpopulation of super-Eddington accreting SMBHs has been identified at large distances~\citep{Maiolino:2023bpi,Suh:2024jbx,2024ApJ...964...39G}. Moreover, the observation of a dormant SMBH at redshift $z=6.68$ suggests that massive BHs are likely assembled through a series of super-Eddigton burst episodes~\citep{2024arXiv240303872J}.

Our analysis is limited to redshifts below $z=6$ and to a typical comoving volume of the Universe, in the sense that we ignore rare peaks in the density field, which may give rise to ultra-luminous quasars~\citep{Nadler:2022kmy}.

Last but not least, we have not included selection effects when comparing with observational data. For this reason, all of our comparisons with observations should be regarded as qualitative.

\subsection{Dependence on hyperparameters}
\label{sec:Dependence_of_hyperparameters}

We do not attempt to perform a comprehensive study of the dependence of our predictions on the model hyperparameters, as this would go beyond the scope of the present work. Therefore, this section focuses on a qualitative comparison between the fiducial results and the effect of varying the main model hyperparameters, one at a time.

\subsubsection{Lower mass threshold for NSC formation}

We have repeated simulations with $M_{\rm cr}=10^6\,M_\odot$, i.e., an order of magnitude smaller than the fiducial value used in our model, fixing all other hyperparameters to their fiducial values.  In this case, the NSCs are typically much lighter (by about an order of magnitude), and not heavier than $10^8\,M_\odot$. As a consequence, massive BH masses are also smaller because the masses of BH seeds strongly depend on the initial conditions of NSCs (cf.~Fig.~\ref{fig:seed_mass_dependence_on_initial_cluster_conditions}).
Moreover, at each galaxy merger episode, an amount of gas proportional to the NSC mass is added to the system (cf.~Sec.~\ref{sec:Major_merger_episodes}). Since, in this case, NSCs are lighter than in the results presented in Sec.~\ref{sec:Results}, BHs accrete less gas as they evolve in the merger tree.
On the other hand, using the fiducial hyperparameters, we have shown that SMBHs can become overmassive relative to the observed local $M_{\rm BH}$--$M_\star$ relation.

\subsubsection{Lower mass ratio threshold for major galaxy mergers}

We have checked the effect of setting the mass ratio threshold to be $Q_{\rm th}=0.01$, instead of the fiducial value of $Q_{\rm th}=0.1$ used in the main text. In this case, the merger rate increases, resulting in more ejections and a smaller density of massive satellite BHs. Nevertheless, this parameter is not critical, in the sense that it does not strongly affect the other observed correlations.

\subsubsection{Smaller gas expulsion time scale}

We have carried out a simulation in which we have set $\tau_{\rm ge}=10\,\rm Myr$, while fixing all other hyperparameters to their fiducial values. Lowering the gas expulsion time scale by an order of magnitude limits the amount of gas accretion onto massive BHs -- their dominant growth channel. Consequently, we find a larger number density of IMBHs compared with the fiducial results, and we find only a few SMBHs in the $(16\,\rm Mpc)^3$ comoving volume. In particular, the heaviest BH in this volume is found to be $\sim10^8\,M_\odot$. In addition, the occupation fraction of SMBHs drops, the slope of the $M_{\rm BH}$--$M_\star$ relation decreases, and we end up underestimating the masses of massive BHs.

\subsubsection{$f_{\rm g,max}=0.1$}

We have repeated our simulations with a lower value of $f_{\rm g,max}=0.1$. This choice limits the amount of gas in NSCs, so we find a larger population of slowly spinning BHs. Massive BHs have lower masses (below $10^8\,M_\odot$) compared to our fiducial model. This is because BHs have less gas to accrete from and therefore they do not grow as much from their seed mass value; it is not due to a lower density of NSCs that can produce heavy seeds. Moreover, the $M_{\rm BH}$--$M_{\star}$ relation in this case is flatter, with a slope of $\simeq0.35$.

\subsubsection{$f_{\star,\rm max}=0.1$}

In the case where we set $f_{\star,\rm max}=0.1$, we limit the initial mass of the formed NSC relative to the host galaxy to be at most $10^6\,M_\odot$. We find a $M_{\rm BH}$--$M_\star$ correlation that agrees more closely with the observed correlation, with a slope of $\simeq1.13$. In this case, massive BHs are lighter because the seed masses are smaller, and not because there is not enough gas accretion.

\section{Conclusions}
\label{sec:Conclusions}

SMBHs are prevalent at the centers of massive galaxies, and their masses scale with galaxy properties (including the velocity dispersion and mass of old stars, and the dark halo mass). Increasing evidence suggests that these trends continue to low stellar masses and towards high redshift.  Seeds are needed for SMBH formation, and this requirement is especially compelling at the highest redshifts explored by JWST. Our study of the hierarchical merging of galaxies over cosmic time, a central tenet of current cosmology, suggests that the central SMBHs found in nearby dwarf spheroidal galaxies may provide evidence for surviving IMBHs  that once were prevalent as SMBH seeds at early epochs.
Our key hypothesis is that such seeds formed in NSCs at early epochs, and we explore this model using cosmological merger trees.

Our model succeeds despite adding ejections. A diverse population of SMBHs emerges, with masses up to (rarely) $\sim 10^{11}M_\odot$. Even if a galaxy loses an SMBH due to a GW kick, a subsequent major merger event with another galaxy usually brings another SMBH into the center, maintaining the high occupation fraction. 

Gas is essential for forming massive BH seeds and for their subsequent growth.
Inefficient star-forming NSCs in which the gas is ejected within a few tens of  Myr do not contribute much to forming massive BH seeds. On the other hand, massive BH seed formation is efficient if the gas is retained longer in the cluster's core, in the sense that in a large region of the parameter space, seeds form with a mass of $\sim10^5M_\odot$, independently of the star formation efficiency.
In total, with our fiducial choice of hyperparameters, we find a massive BH mass of $\approx1.7\times10^{11}\,M_\odot$ (not including stellar-mass BHs) within the $(16\,\rm Mpc)^3$ cosmological box. This translates to $\approx4.2\times10^{16}\,M_\odot\,\rm Gpc^{-3}$ which is a factor of $\approx10^{-4}$ less than the critical density to close the Universe. For comparison, according to \cite{1982MNRAS.200..115S}, the total mass accumulated by quasars is at least $4.7\times10^{13}\,M_\odot\,\rm Gpc^{-3}$, where the accretion efficiency is set to $10\%$.

Notice that a galaxy that has lost its nuclear SMBH may regain it if it merges with another galaxy that hosts a central SMBH, provided that it has a rich merger history. 
Therefore, galaxies that have undergone several major merger episodes will likely contain nuclear massive BHs, and a few wandering off-nuclear massive BHs. 
The off-nuclear BHs could either be surrounded by a stellar system (classified as an ultra-compact dwarf), indicating a minor merger episode, or completely isolated, if they received a large enough GW kick. 
Such isolated SMBHs could only be identified either by their microlensing effect on foreground stars, or by their electromagnetic signatures when accreting from the interstellar medium. 

Our code has many free parameters, that we have chosen to give a crude fit to local scaling laws involving  NSCs.  
We caution the reader that this is in the same spirit as other discussions of semi-analytical models of galaxy formation, e.g., \cite{Nadler:2022kmy}. Independent studies using different prescriptions are needed to test the robustness of our results. Our code is publicly available on {\sc GitHub}: \url{https://github.com/Kkritos/nsc-tree}.

We anticipate future applications that will include the use of  
faster codes capable of generating merger tree histories for rare halos hosting extremely luminous high-redshift galaxies (as seen by JWST),  
follow cosmological simulations of halo assembly history from $z>10$ down to mass scales $\sim 10^5\,M_\odot, $ and incorporate dynamical modeling of NSCs.

We have also predicted GW merger rates, including massive BH binaries, extreme-mass ratio inspirals, and stellar-mass BH binaries. Future experiments such as LISA, the Einstein Telescope, and Cosmic Explorer could detect these sources, either as a stochastic GW background or through the observation of individual merger events.

\begin{acknowledgments}
We thank Francesco Iacovelli and Luca Reali for discussions.
K.K., E.B., and S.Y. are supported by NSF Grants No. AST-2307146, No. PHY2207502, No. PHY-090003, and No. PHY-20043, by NASA Grants No. 20-LPS20-0011 and No. 21-ATP21-0010, by the John Templeton Foundation Grant 62840, by the Simons Foundation, and by the Italian Ministry of Foreign Affairs and International Cooperation Grant No. PGR01167.
K.K. is supported by the Onassis Foundation - Scholarship ID: F ZT 041-1/2023-2024.
S.Y. is supported by the NSF Graduate Research Fellowship Program under Grant No. DGE2139757.
This work was carried out at the Advanced Research Computing at Hopkins (ARCH) core facility (\url{rockfish.jhu.edu}), which is supported by the NSF Grant No.~OAC-1920103.
For the purpose of open access, the author has applied a Creative Commons Attribution (CC BY) license to any Author Accepted Manuscript version arising from this submission.
\end{acknowledgments}

\bibliographystyle{aasjournal}
\bibliography{merger-trees}

\end{document}